\documentclass[]{interact}

\usepackage{epstopdf}
\usepackage{commath}
\usepackage{amsfonts}
\usepackage[table]{xcolor}
\usepackage{amsthm}
\usepackage{enumerate}
\usepackage{enumitem}
\usepackage{geometry}
\usepackage{mathtools}
\usepackage{bbold}
\usepackage{parskip}
\usepackage{float}
\usepackage{caption}
\usepackage{subcaption}
\usepackage{threeparttable}
\usepackage{booktabs}
\usepackage[pdftex,colorlinks=true]{hyperref}

\usepackage{todonotes}

\usepackage[numbers,sort&compress]{natbib}

\bibpunct[, ]{[}{]}{,}{n}{,}{,}

\newcommand{\ee}[0]{\ensuremath{\xi_+}}
\newcommand{\rzero}[0]{\ensuremath{\mathcal{R}_{0}}}
\newcommand{\Seq}[0]{\ensuremath{S^{*}}}
\newcommand{\Ieq}[0]{\ensuremath{I^{*}}}
\newcommand{\Req}[0]{\ensuremath{R^{*}}}

\newcommand{\deriv}[0]{\ensuremath{\mathrm{d}}}
\newcommand{\taylor}[2]{\left(#1\right)_{#2}}
\newcommand{\reals}[0]{\ensuremath{\mathbb{R}}}
\newcommand{\domain}[0]{\ensuremath{\mathcal{D}}}
\newcommand{\matcont}[0]{\texttt{MatCont}}
\newcommand{\gh}[0]{\ensuremath{\mathrm{GH}}}
\newcommand{\h}[0]{\ensuremath{\mathrm{H}}}
\newcommand{\xvec}[0]{\ensuremath{\mathbf{x}}}
\renewcommand\bibfont{\fontsize{10}{12}\selectfont}

\theoremstyle{plain}
\newtheorem{theorem}{Theorem}
\newtheorem{lemma}{Lemma}

\theoremstyle{definition}

\theoremstyle{remark}

\begin{document}


\title{Dynamics of an 
SIRWS model with waning of immunity and  varying immune boosting period}

\author{
\name{Richmond Opoku-Sarkodie\textsuperscript{a}\thanks{CONTACT Ferenc A. Bartha. Email: barfer@math.u-szeged.hu}, Ferenc A. Bartha \textsuperscript{b}, M\'{o}nika Polner \textsuperscript{c}, and  Gergely  R\"ost\textsuperscript{d}  }
\affil{\textsuperscript{a,}\textsuperscript{b,}\textsuperscript{c,}\textsuperscript{d} Bolyai Institute, University of Szeged, H-6720 Szeged, Aradi v\'ertan\'uk tere 1, Hungary\\
E-mail: ropokusarkodie@gmail.com \textsuperscript{a};
 barfer@math.u-szeged.hu \textsuperscript{b};\\
 polner@math.u-szeged.hu \textsuperscript{c}; rost@math.u-szeged.hu \textsuperscript{d};
 }
}

\maketitle

\begin{abstract}
SIRS models capture transmission dynamics of infectious diseases for which immunity is not lifelong. Extending these models by a $W$ compartment for individuals with waning immunity, the boosting of the immune system upon repeated exposure may be incorporated. Previous analyses assumed identical waning rates from $R$ to $W$ and from $W$ to $S$. This implicitly assumes equal length for the period of full immunity and of waned immunity. We relax this restriction, and allow an asymmetric partitioning of the total immune period. Stability switches of the endemic equilibrium are investigated with a combination of analytic and numerical tools. Then, continuation methods are applied to track bifurcations along the equilibrium branch. We find rich dynamics: Hopf bifurcations, endemic double bubbles, and regions of bistability. Our results highlight that the length of the period in which waning immunity can be boosted is a crucial parameter significantly influencing long term epidemiological dynamics.
\end{abstract}


\section{Introduction}
The susceptible-infectious-recovered (SIR) approach has been widely applied in diverse forms to understand the transmission dynamics of communicable diseases. For many infections, immunity is not lifelong, and after some time, recovered individuals may become susceptible again. Prior to that, repeated exposure to the pathogen might boost the immune system, thus prolonging the length of immune period. A very general framework of waning-boosting dynamics has been introduced in \cite{general}.
Special cases of that are the SIRWS compartmentals models, where $W$ is the collection of individuals whose immunity is waning but can be boosted upon repeated exposure without experiencing the disease again.

SIRWS models formulated as systems of ordinary differential equations were studied in  \cite{dafilis2012influence,lavine2011natural,leung2016periodic}. In these models, the immunity period  is divided into two parts: upon recovery, previously infected individuals move to $R$, and from there they may transit to $W$ as time elapses. If they are exposed again while being in $W$, their immunity can be boosted and they move back $R$. Otherwise, they eventually lose their immunity, become susceptible again an move back to $S$. The aforementioned studies model these two phases of the immune period by a symmetric partitioning, by
assuming identical rates of transition from $R$ to $W$ and from $W$ to $S$. 

In contrast, our work removes this symmetry constraint, and we analyze how the different partitioning of the immune period into $R$ and $W$, and varying boosting rates affect the dynamics of the model. First, the existence of equilibria and analytic conditions for their 
local stability are established. Then, using numerical tools and methods, we observe the emergence of complex phenomena through various bifurcations, such as endemic double bubbles, 
and multiple regions of bistability.

\subsection{Modified SIRWS model}\label{class}
\begin{figure}[!ht]
	\centering
	\includegraphics[width=0.8\textwidth]{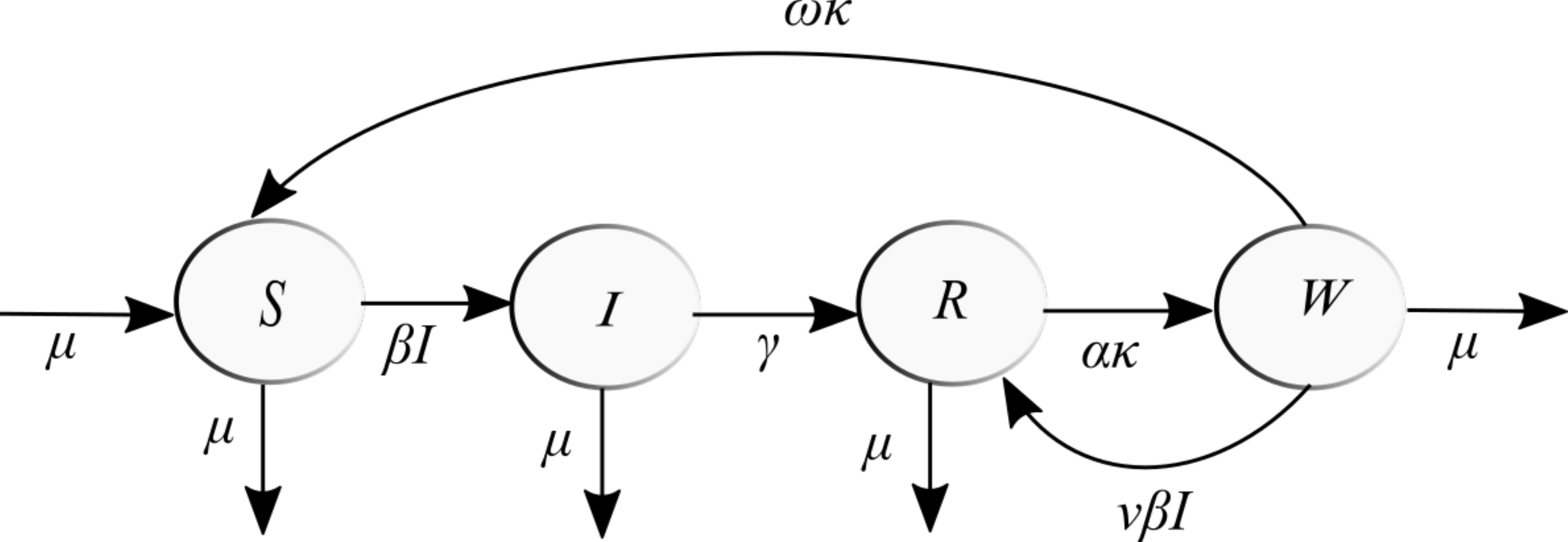}
	\caption{Flow diagram of the SIRWS system \eqref{eq:sirw}.}\label{diagramSIRWS}
\end{figure}

This section describes the SIRWS compartmental model in which the 
population is divided as follows. 
The individuals susceptible to infection are placed in $S$, 
those currently infectious in $I$, and those recovered from infection 
are divided into two compartments based on their immunity level.
The fully immune are found in $R$ and those with waned immunity are in $W$. 
Figure~\ref{diagramSIRWS} depicts the flow diagram of our model 
governed by the system of ordinary differential equations
\begin{subequations}\label{eq:sirw}
	\begin{align} 
	\frac{dS}{dt} & =
	    \, -\beta I S + \omega \kappa W + \mu (1 - S), \label{eq:sirw_0}\\
	\frac{dI}{dt} & =
	    \, \beta I S - \gamma I - \mu I, \label{eq:sirw_1}\\
	\frac{dR}{dt} & =
	    \, \gamma I - \alpha \kappa R + \nu \beta I W - \mu R, \label{eq:sirw_2}\\
	\frac{dW}{dt} & =
	    \, \alpha \kappa R - \omega \kappa W - \nu \beta I W - \mu W, \label{eq:sirw_3}
	\end{align}
\end{subequations}
where $\beta$, $\gamma$, and $\mu$ are referred to as 
the infection rate, recovery rate, and birth and death rate, respectively. 

Recovered individuals may lose immunity by the chain of transitions 
$R \to W \to S$. The average duration of immune protection, that is the 
average time required to complete both of these transitions is $\kappa^{-1}$
and, hence, $\kappa$ is the immunity waning rate.
Members of $W$ are still immune to infection and are subject to immune boosting 
upon re-exposure, the frequency of which is modulated by the 
boosting force $\nu$. In our analysis, 
hosts going through boosting are not infectious, such as in \cite{general,dafilis2012influence,lavine2011natural,leung2018infection}, as opposed to \cite{strube}.

The population is normalized to $1$ 
that is $N(t) = S(t) + I(t) + R(t) + W(t) = 1$ for all $t\geq 0$. 
Vital dynamics is modeled with the rate $\mu$ for birth and death, and disease induced fatality is not considered. 

In former SIRWS model studies, e.g. \cite{dafilis2012influence,lavine2011natural,leung2018infection}, 
the immune waning rates are the same for individuals 
who move from the recovered compartment to the waning compartment and 
for those who transition onward to the susceptible compartment. 
In contrast, we consider an asymmetric partition of the immunity period by 
introducing the parameters $\alpha > 1$ and $\omega > 1$ setting 
the average time spent in $R$ and $W$ to 
$(\alpha \kappa)^{-1}$ and $(\omega \kappa)^{-1}$, 
respectively. Hence,
\begin{equation}
\label{eq:alpha-omega}
    \frac{1}{\alpha \kappa} + \frac{1}{\omega \kappa} = \frac{1}{\kappa} 
    \qquad \mbox{that is} \qquad
    \omega = \frac{\alpha}{\alpha - 1}.
\end{equation} 
Note that the special case $\alpha = \omega = 2$, representing 
the symmetric partition of immunity period, is what was 
considered in the aforementioned studies. 

By considering various limiting scenarios of boosting for \eqref{eq:sirw}, it is 
apparent that the system exhibits 
SIRS-like dynamics as $\nu \rightarrow 0^+$ and 
SIR-like dynamics as $\nu \rightarrow \infty$. 
In addition, we observe
SIR-like dynamics as $\alpha \rightarrow 1^{+}$ ~ ($\omega \rightarrow \infty$)  
and SIS-like dynamics as $\alpha \rightarrow \infty$ ~ ($\omega\rightarrow 1^{+}$).

\section{Equilibria and stability}
\label{sec:EE-analytic}
This section first investigates system \eqref{eq:sirw} in order to establish the 
formulae for the equilibria of our SIRWS model. Then, 
we analyze the transcritical bifurcation where these equilibria exchange stability 
in Section~\ref{sec:transcritical}. Finally, we derive the Routh-Hurwitz stability 
criterion in Section~\ref{sec:LAS-EE}.

We begin by utilizing the relation 
$$W(t) = 1 - S(t) - I(t) - R(t),$$
to obtain the reduced system 
\begin{subequations}\label{eq:reduced}
	\begin{align} 
	\frac{dS}{dt} &=\, 
	    -\beta I S + \omega \kappa (1 - S - I - R) + \mu(1 - S), 
	        \label{eq:te_0} \\
	\frac{dI}{dt} &=\, 
	    \beta I S - \gamma I - \mu I, 
	        \label{eq:te_1} \\
	\frac{dR}{dt} &=\, 
	    \gamma I - \alpha \kappa R + \nu \beta I(1 - S - I - R) - \mu R. 
	        \label{eq:te_2}
	\end{align}
\end{subequations}
Note that the region relevant for our epidemiological setting 
\begin{equation*}
    (S(t), I(t), R(t)) \in \domain := 
    \left\{ (s, i, r) \in \reals_{\geq 0}^3 ~ | ~ 0 \leq s + i + r \leq 1 \right\} 
\end{equation*}
is forward invariant. 

\vspace{1em}

Now, let us turn our attention to equilibria of \eqref{eq:reduced} 
and seek solutions of the steady state equations
\begin{subequations}\label{eq:equilib}
\begin{align} 
	- \beta \Ieq \Seq + \omega \kappa (1 - \Seq - \Ieq - \Req) + 
	    \mu (1 - \Seq) &= 0, 
	    \label{eq:equilib_0}\\
	\beta \Ieq \Seq - \gamma \Ieq - \mu \Ieq &= 0 ,
	    \label{eq:equilib_1}\\
	\gamma \Ieq - \alpha \kappa \Req + \nu \beta 
	    \Ieq(1 - \Seq - \Ieq - \Req) - \mu \Req &= 0. 
	    \label{eq:equilib_2}
\end{align}
\end{subequations}

From \eqref{eq:equilib_1}, we obtain that either $\Ieq = 0$ or 
$\Seq = \tfrac{\gamma + \mu}{\beta}$. In the first case, 
$\Req = 0$ follows from \eqref{eq:equilib_2} 
and, finally, $\Seq = 1$ from \eqref{eq:equilib_0}. 
Hence we obtain 
\begin{equation*}
    \xi_0 = (1, 0, 0), 
\end{equation*} 
the \emph{disease free equilibrium} (DFE) of \eqref{eq:reduced}. 
In the latter case when 
\begin{equation}
\label{eq:Seq}
    \Seq = \tfrac{\gamma + \mu}{\beta}, 
\end{equation}
equation \eqref{eq:equilib_0} yields 
\begin{equation*}
\begin{split}
    \Ieq 
    = \frac{(\mu + \omega \kappa) (1 - \Seq) - 
        \omega \kappa \Req}{\beta \Seq + \omega \kappa} 
    &= \frac{c_0 c_1}{\beta} - 
        \frac{\omega \kappa}{\gamma + \mu + \omega \kappa} \Req,
\end{split}
\end{equation*}
with 
\begin{equation*}
\begin{split}
    c_0 &=
        \frac{1}{\gamma + \mu + \omega \kappa} \cdot 
        \left( 1 + \frac{\omega \kappa}{\mu} \right)
        \qquad \mbox{and} \\
    c_1 &= \mu (\beta - (\gamma + \mu)). \\
\end{split}
\end{equation*}

Then, using the formulae for $\Seq$ and $\Ieq$, 
\eqref{eq:equilib_2} results in a quadratic equation for $\Req$. 
It is straightforward to verify that the leading term coefficient is 
positive, hence, the graph of it is an open up parabola 
with the $y$-intercept  
\begin{equation*}
    \frac{\gamma c_0 c_1}{\beta} \left( 
        1 + \frac{\nu c_0 c_1}{\mu + \omega \kappa} 
    \right).
\end{equation*}
Moreover, as shown in Appendix~\ref{app:Rstar}, 
the solutions can be expressed as
\begin{equation}
\label{eq:Req}
\begin{split}
    \Req_{\pm} 
    &= \frac{\gamma + \mu + \omega \kappa}{2 \beta \omega \kappa} \left[ 
    \left( 
        2 c_0 - 
        \frac{1}{\gamma + \mu}
    \right) c_1 + 
    \frac{1}{
        \nu (\gamma + \mu)}
    \left(
        c_2 \mp \sqrt{(c_1 \nu + c_2)^2 + c_3 \nu}
    \right) \right],
\end{split}
\end{equation}
using $c_2, c_3$ given by  
\begin{equation*}
\begin{split}
    c_2 &= (\gamma + \mu) (\alpha \kappa + \omega \kappa) + 
        \mu (\gamma + \mu) + \alpha \omega \kappa^2 
        \qquad \mbox{and} \\
    c_3 &= 4 \gamma (\beta - (\gamma + \mu)) \alpha \omega \kappa^2.
\end{split}
\end{equation*}
Finally, substituting \eqref{eq:Req} into the formula for $\Ieq$ results in 

\begin{equation}
\label{eq:Ieq}
    \Ieq_{\pm} = 
        \frac{\pm \sqrt{(c_1 \nu + c_2)^2 + c_3 \nu} + (c_1 \nu - c_2)}{
        2 \beta \nu (\gamma + \mu)}.
\end{equation}

Hence, we obtained the two remaining equilibria of 
\eqref{eq:reduced}, namely the {\emph{endemic equilibrium} (EE)}
\begin{equation*}
\begin{split}
    \xi_{+} &= (\Seq, \Ieq_{+}, \Req_{+}),  \\
    ~&~ \qquad \mbox{and} \\
    \xi_{-} &= (\Seq, \Ieq_{-}, \Req_{-}).
\end{split}
\end{equation*}

\vspace{0.5em}

Clearly, $\Ieq_{-} \leq 0$ whenever the square root is real as 
the inequality is readily satisfied for $\beta < \gamma + \mu$ 
(then $c_1, c_3 < 0$) and directly follows from 
\begin{equation}
\label{eq:sqrt-c-eq}
    \sqrt{(c_1 \nu +c_2)^2 + c_3 \nu} \geq |c_1 \nu - c_2| 
    \quad 
    \Leftrightarrow \quad 4 c_1 c_2 + c_3 \geq 0,
\end{equation} 
when $\beta \geq \gamma + \mu$ (then $c_1, c_3 \geq 0$). 
Moreover, the condition $\beta \geq \gamma + \mu$ is sufficient 
(but not necessary) for $\Ieq_{\pm} \in \reals$.
Obviously, in the epidemiological setting of this manuscript, 
solely $\xi_{+}$ may be admissible.

Another important implication of \eqref{eq:sqrt-c-eq} is that
\begin{equation*}
    \Ieq_{+} > 0 \Leftrightarrow \beta > \gamma + \mu
    \quad \mbox{and} \quad 
    \Ieq_{+} = 0 ~ \mbox{for} ~ \beta = \gamma + \mu. 
\end{equation*} 
Observe that, in the case of equality, $\xi_0 = \xi_+$ holds.  
Furthermore, again for $\beta \geq \gamma + \mu$, 
the parabola for $R^*$ has a positive $y$-intercept,
thus, both solutions are either positive or negative.  
Moreover, we have 
$\Req_{-} > 0$ as 
$2 c_0 - \tfrac{1}{\gamma + \mu} > 0$ is satisfied and 
$c_1 \geq 0$. These imply the positivity of the other root that is 
$\Req_{+} > 0$. 

Now, summing 
\eqref{eq:equilib_0}, \eqref{eq:equilib_1}, and \eqref{eq:equilib_2} 
results in 
\begin{equation*}
    (\omega \kappa + \mu + \nu \beta \Ieq)(1 - \Seq - \Ieq - \Req) - 
    \alpha \kappa \Req = 0,
\end{equation*}
hence, $\Seq + \Ieq + \Req \leq 1$ must hold for 
non-negative $\Seq, \Ieq, \Req$
implying 
$\xi_{+} \in \domain \Leftrightarrow \beta \geq \gamma + \mu$. 

\vspace{0.5em}

Finally, note that the \emph{basic reproduction number}, 
see {\it e.g.} \cite{general}, 
of the system 
\eqref{eq:reduced} -- and of \eqref{eq:sirw} -- is 
\begin{equation*}
    \rzero = \frac{\beta}{\gamma + \mu},  
\end{equation*}
thus, we may rewrite the condition $\beta \geq \gamma + \mu$ as 
$\rzero \geq 1$.

\vspace{1em}

Before continuing the analysis, 
let us summarize our findings so far.
\begin{itemize}
    \item There is a unique DFE $\xi_0 \in \domain$, 
        which exists for all parameter values in the system.
    \item If $R_0 \leq 1$, then there is no other equilibrium in $\domain$.
    \item If $R_0 > 1$, then there is a unique, 
        positive EE $\ee \in \domain$.
\end{itemize}

\subsection{Transcritical bifurcation at \texorpdfstring{$\rzero=1$}{R0 = 1}}
\label{sec:transcritical}
For the stability analysis of the disease free equilibrium $\xi_0$, 
consider the Jacobian matrix for our SIRWS system \eqref{eq:reduced}
\begin{equation}\label{eq:jacobian}
J=\left[\begin{array}{ccc}
    -(\omega\kappa+\mu+\beta I) & -\beta S-\omega\kappa & -\omega\kappa \\
    \beta I & -(\gamma+\mu-\beta S) & 0 \\
    -\nu\beta I & \gamma-2\nu\beta I+\nu\beta(1-S-R) & -(\alpha\kappa+\mu+\nu\beta I) 
  \end{array}\right]
\end{equation}
and evaluate at the DFE $\xi_0$ 
\[
J|_{\xi_0}=\left[\begin{array}{ccc}
     -(\omega\kappa+\mu) & -\beta -\omega\kappa & -\omega\kappa \\
    0 & -(\gamma+\mu-\beta ) & 0 \\
    0 & \gamma & -(\alpha\kappa+\mu) 
  \end{array}\right].
\]
Then, the corresponding eigenvalues are
\[
\quad \lambda_{1}=\beta-(\gamma+\mu),
\quad \lambda_{2}=-(\mu+\alpha\kappa), 
\quad \mbox{and} \quad \lambda_{3}=-(\mu+\omega\kappa).
\]
The two eigenvalues $\lambda_2, \lambda_3$ are negative and 
$\lambda_1 < 0$ iff $\beta < \gamma + \mu$. 
Hence, the DFE is locally asymptotically stable when $\rzero < 1$ and 
unstable for $\rzero > 1$.

\begin{figure}[!ht]
	\centering
	 \includegraphics[scale=0.7]{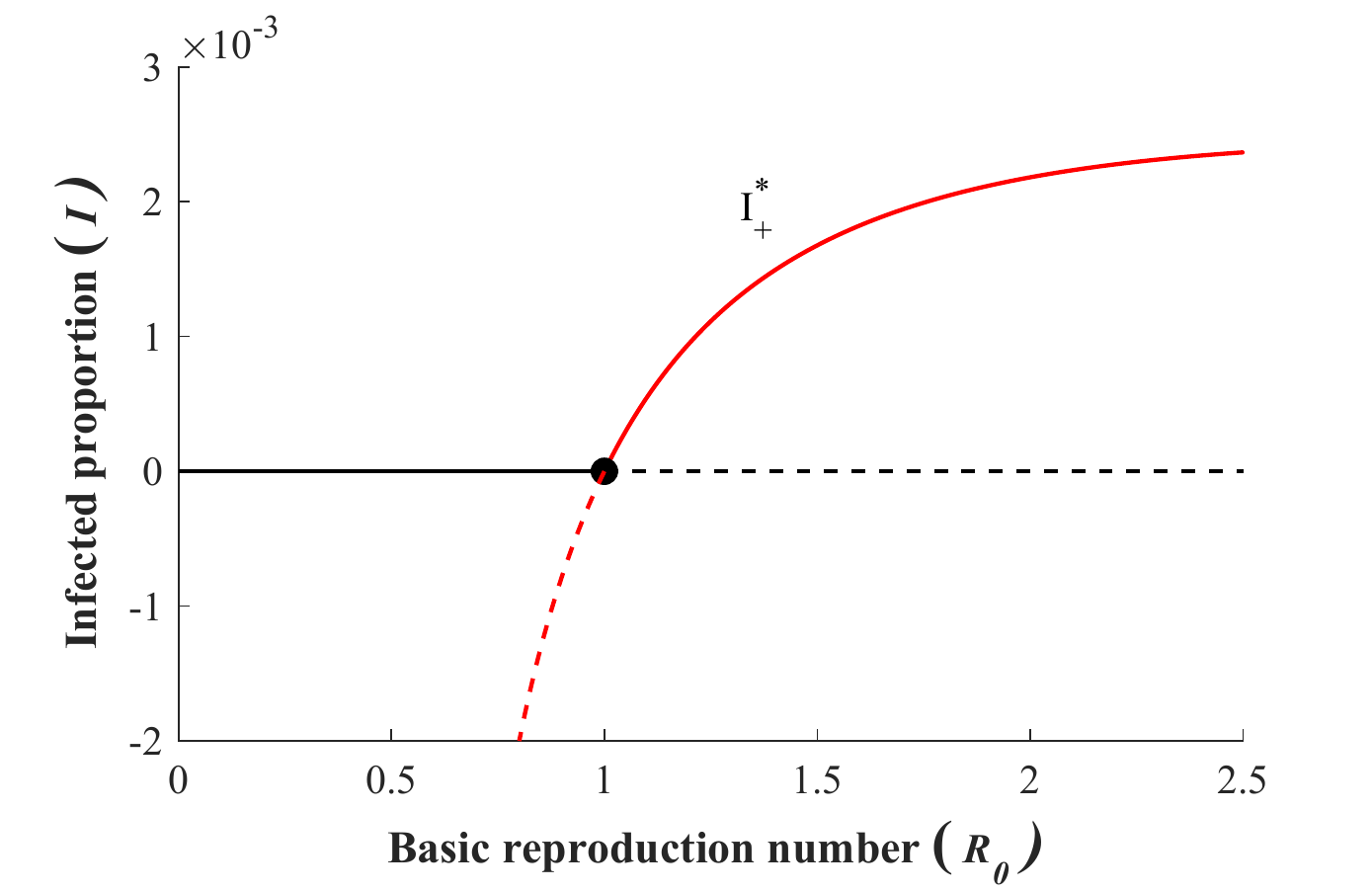}
	\caption{
	    Transcritical bifurcation of forward type and 
	    the appearance of the LAS endemic equilibrium $\ee$ at $\rzero = 1$.
	}\label{fig:transcritical}
\end{figure}

The following Theorem describes the bifurcation associated with this 
stability change at $\rzero = 1$ that is 
also demonstrated in Figure~\ref{fig:transcritical}. 
The proof relies on Theorem~4.1 of \cite{castillo2004dynamical}  
based on center manifold theory \cite{carr2006center, wiggins2013global}. 
For the sake of completeness, the relevant version of the original theorem is included 
in Appendix~\ref{app:transcritical}. 

\begin{theorem}
\label{thm:bifurcation}
    A transcritical bifurcation of forward-type occurs at $\rzero = 1$. 
\end{theorem}

\begin{proof}
Fix all parameters but $\beta$ that will serve as the 
bifurcation parameter with $\beta^{*} = \gamma + \mu$ 
corresponding to the critical case $\rzero = 1$.

We show that the conditions of Theorem~\ref{app:thm-transcritical} 
are satisfied for the system $\dot{\xvec} = f(\xvec, b)$, 
where 
$$f = (f_1, f_2, f_3) \equiv (f_{S}, f_{I}, f_{R})$$ 
is obtained by applying the substitutions 
$\beta \to b + \beta^*$ and 
$(S, I, R) \to (x_S, x_I, x_R) + \xi_0$ to equations 
\eqref{eq:te_0}, \eqref{eq:te_1}, and \eqref{eq:te_2}, 
with 
$$\xvec = (x_1, x_2, x_3) \equiv 
(x_S, x_I, x_R).$$  

\vspace{2em}

The matrix $A = D_\xvec f(\mathbf{0}, 0)$ 
($ = J|\xi_0$ with $\beta = \beta^*$) 
has one simple zero eigenvalue and 
two eigenvalues with negative real part 
\[
    \lambda_1 = 0, \quad
    \lambda_2 = -(\mu + \alpha \kappa), \quad 
    \lambda_3 = -(\mu + \omega \kappa).
\]

Now, let us calculate 
\begin{align*}
    Z_1 &= \sum_{k, i, j = 1}^{3} v_{k} w_{i} w_{j} 
        \frac{\partial^{2} f_{k}}{\partial x_{i} \partial x_{j}}
            (\mathbf{0}, 0) \qquad \mbox{and} \\ 
    Z_2 &= \sum_{k, i = 1}^{3} v_{k} w_{i} 
        \frac{\partial^{2} f_{k}}{\partial x_{i} \partial b}
            (\mathbf{0}, 0),
\end{align*}
where $w, v$ are the right and left eigenvectors of $A$ 
corresponding to the zero eigenvalue. 

Note that we may fix $w_{2} = 1$ as $A w = 0$ is underdetermined. 
Then, 
\begin{align*}
    w_{1} &= 
        - \left[1 + \frac{\alpha \kappa \gamma}{(\omega \kappa + \mu)
            (\alpha \kappa + \mu)}
        + \frac{\gamma}{\alpha \kappa + \mu} \right] \qquad \mbox{and} \\
    w_{3} &= \frac{\gamma}{\alpha \kappa + \mu}
\end{align*}
follow. Analogously, we find a left eigenvector $v = (0, 1, 0)$. 

As $v_{1} = v_{3} = 0$, the sums get reduced to the terms containing 
$$f_2 \equiv f_I = (b + \beta^*) x_I (x_S + 1) - 
(\gamma + \mu) x_I.$$
Clearly, the nonzero second order partial derivatives of 
$f_{I}$ at $(\mathbf{0}, 0)$ are 
\[
    \frac{\partial^{2} f_{I}}{\partial x_I \partial b}
        (\mathbf{0}, 0) = 1 \quad \mbox{and} \quad 
    \frac{\partial^{2} f_{I}}{\partial x_I \partial x_S}
        (\mathbf{0}, 0) = \beta^{*} = \gamma + \mu.
\]
\
Hence,
\begin{align*}
    Z_1 &= 2 v_{2} w_{1} w_{2} \frac{\partial^{2} f_{I}}{\partial x_I \partial x_S} 
        (\mathbf{0}, 0) \\
        &= -2 \left[1 + \frac{\alpha \kappa \gamma}{(\omega \kappa + \mu)
            (\alpha \kappa + \mu)}
        + \frac{\gamma}{\alpha \kappa + \mu} \right] (\gamma + \mu)
        \quad \mbox{and} \\
    Z_2 &= v_{2} w_{2} 
        \frac{\partial^{2} f_{I}}{\partial x_I \partial b}
        (\mathbf{0}, 0) = 1. 
\end{align*}

As $Z_1 < 0$ and $Z_2 > 0$ for all parameters, 
we can apply Theorem~\ref{app:thm-transcritical}
noting that even though $w_1 < 0$, as 
the first component of $\xi_0 = (1, 0, 0)$ is positive, 
$w_1 \geq 0$ is not required actually. 

Translating the statement of the aforementioned Theorem to 
our original system \eqref{eq:reduced}, we obtain that 
when $\rzero$ increases through 1, 
a transcritical bifurcation of forward type occurs with 
$\xi_0$ losing and $\ee$ gaining local asymptotic stability (LAS), respectively. 
\end{proof}


\subsection{The Routh-Hurwitz criterion for \texorpdfstring{$\ee$}{the endemic equilibrium}}
\label{sec:LAS-EE}
This section analyzes the stability of the 
endemic equilibrium $\ee$ for fixed $\beta, \gamma, \kappa$, and $\mu$, given that  $\rzero > 1$ holds. 

Local asymptotic stability (LAS) is characterized by all eigenvalues 
of the Jacobian \eqref{eq:jacobian} at $\ee$ having negative real part. 
Therefore, we consider the matrix
\begin{equation*}
J|_{\ee} = 
  \left[ \begin{array}{cccc}
    - (\omega \kappa + \mu + \beta \Ieq_+) & - (\gamma + \mu + \omega \kappa) 
        & -\omega\kappa \\
    \beta \Ieq_+ & 0 & 0 \\
    - \nu \beta \Ieq_+ & \gamma - 2 \nu \beta \Ieq_+ + \nu \beta (1 - \Seq - \Req_+) 
        & - (\alpha \kappa + \mu + \nu \beta \Ieq_+) 
  \end{array} \right] 
\end{equation*}
and, in turn, its characteristic polynomial 
\begin{equation*}
  a_0 \lambda^3 + a_1 \lambda^2 + a_2 \lambda + a_3 = 0,
\end{equation*}
with 
\begin{equation}
\begin{split}
\label{eq:RH-coeffs}
  a_0 &= 1, \\
  a_1 &= \beta \Ieq_+ (1 + \nu) + (\alpha \kappa + \omega \kappa + 2 \mu), \\
  a_2 &= \beta \Ieq_+ [ (\alpha \kappa + \omega \kappa + 2 \mu) + 
    \gamma + \beta \nu \Ieq_+ + \mu \nu ] + 
    (\omega \kappa + \mu) (\alpha \kappa + \mu), \\
  a_3 &= \beta \Ieq_+ [ (\omega \kappa + \mu) (\alpha \kappa + \mu) + 
    (\gamma + \mu) \beta \nu \Ieq_+ + 
    \gamma (\alpha \kappa + \omega \kappa + \mu) ~ + \\
    & ~ ~ ~ ~ \omega \kappa \beta \nu (1 - \Seq - \Ieq_+ - \Req_+) ],  
\end{split}
\end{equation}
and $\Seq, \Ieq_+, \Req_+$ as given in \eqref{eq:Seq}, \eqref{eq:Req}, and 
\eqref{eq:Ieq}. 

Utilizing the Routh-Hurwitz (RH) criterion \cite{routh1877treatise,murray2007mathematical} 
yields that $\ee$ is LAS iff the following inequalities 
are satisfied
\begin{equation*}
\begin{split}
    a_i &> 0, \quad \mbox{for } i = 0, 1, 2, 3, ~ \mbox{and} \\
    a_1 a_2 &> a_3.
\end{split}
\end{equation*}
As the positivity of $a_0, \ldots, a_3$ is trivial, 
we are led to analyze the sign changes of the function
\begin{equation}\label{eq:RH}
  y_{\nu}(\alpha)=a_1 a_2 - a_3,
\end{equation}
for $\alpha > 1$ and $\nu > 0$. 

\subsubsection{Transformation of \texorpdfstring{$y_{\nu}(\alpha)$}{the Routh-Hurwitz criterion}}
\label{sec:LAS-EE-Eta}

The formulae in \eqref{eq:Ieq} and \eqref{eq:RH-coeffs}  
appear to be (mostly) symmetric with respect to $\alpha$ and $\omega$. 
Recall that 
these two parameters are closely related as
\begin{equation*}
    \alpha + \omega = \alpha \omega = \frac{\alpha^2}{\alpha - 1} 
\end{equation*}
directly follows from \eqref{eq:alpha-omega}. These considerations suggest to 
introduce the substitution 
\begin{equation}
\label{eq:eta}
    \eta = \kappa(\alpha + \omega) = \kappa(\alpha \omega) = 
    \kappa \frac{\alpha^2}{\alpha - 1}, 
\end{equation}
with $\eta \in [4 \kappa, \infty)$ and the 
$\alpha = \omega = 2$ case corresponding to 
$\eta = 4 \kappa$. 
Nevertheless, in order to apply \eqref{eq:eta}, 
we need to establish that 
$a_3$ in \eqref{eq:RH-coeffs} may be considered as a function of $\eta$. This holds due to the equality 
\begin{equation*}
    \omega \kappa \beta \nu (1 - \Seq - \Ieq_+ - \Req_+) = 
    \beta \nu (\gamma + \mu) \Ieq_+ - c_1 \nu, 
\end{equation*}
see Appendix~\ref{app:RH-EE-Eta} for details. 

Then, one obtains that
\begin{equation*}
    y_{\nu}(\alpha) \equiv y_{\nu}(\eta) = \hat a_1 \hat a_2 - \hat a_3,
\end{equation*}
with 
\begin{equation*}
\begin{split}
    \hat a_1 &= \hat I (1 + \nu) + (\eta + 2 \mu), \\
    \hat a_2 &= \hat I [ (\eta + \mu) + (\gamma + \mu) + \mu \nu + 
        \nu \hat I ] + \mu (\eta + \mu) + \kappa \eta, \\
    \hat a_3 &= \hat I [ 2 \nu \hat I(\gamma + \mu)  
        - \nu \mu (\beta - (\gamma + \mu)) + 
        (\gamma + \mu) (\mu + \eta) + \kappa \eta ], 
\end{split}
\end{equation*}
where $\hat I = \beta \Ieq_+$. 

Substitution \eqref{eq:eta} reveals an important 
feature of $y_{\nu}(\alpha)$, namely, there is a bijection  
$(1, 2) \ni \alpha \mapsto \alpha' \in (2, \infty)$ 
such that $y_{\nu}(\alpha) = y_{\nu}(\alpha')$.  
In particular, local extrema at $\alpha \neq 2$ appear in pairs.

Furthermore, using the chain rule, we obtain that 
\begin{equation*}
    \frac{\partial y_\nu}{\partial \alpha} = 
    \frac{\partial y_\nu}{\partial \eta} \cdot \frac{\deriv \eta} {\deriv \alpha} = 
    \frac{\partial y_\nu}{\partial \eta} \cdot 
        \frac{\kappa \alpha (\alpha - 2)}{(\alpha - 1)^2}. 
\end{equation*}
Clearly, $\alpha = 2$ (that is $\eta = 4 \kappa$) is a critical point of $y_\nu$ for 
all immune boosting parameters $\nu$. By Lemma~\ref{app:lemma-composite-zeros}, 
either all derivatives of $y_\nu$ are zero at $\alpha = 2$ or the first 
non-vanishing derivative is of even order. As $y_\nu$ is analytic 
and not identically zero for any 
$\rzero > 1$, the former is not possible, hence, $\alpha = 2$ is a 
local extremum for all boosting rates $\nu$.

\section{Numerical analysis}
\label{sec:numerical-analysis}

This section summarizes the results 
of our numerical stability and bifurcation analysis of system \eqref{eq:reduced}
with respect to varying waning and boosting dynamics. 
In the remaining part of the manuscript, all other parameters are considered 
to be fixed following \cite{lavine2011natural} to model pertussis as 
\begin{equation}
\label{eq:parametrization}
\begin{split}
    &\gamma = 17, \\
    &\kappa = 1/10, \\
    &\mu = 1/80, \\
    &\beta=260, \\
\end{split}
\end{equation}
corresponding to an average infectious period of 21 days, average life expectancy of 80 years
and a basic reproduction number $\rzero = 15.28$.

First, Section~\ref{sec:LAS-EE-numerics} discusses how the local stability of 
$\ee$ changes given \eqref{eq:parametrization} with varying $\alpha$ and $\nu$. Then,
Section~\ref{sec:bif-alpha} analyzes these stability changes and the corresponding bifurcations. 
In addition, using numerical continuation methods, we observe the 
bistable regions in the $(\alpha, \nu)$-plane.

\subsection{Analysis of the Routh-Hurwitz criterion for  \texorpdfstring{$\ee$}{the endemic equilibrium}}
\label{sec:LAS-EE-numerics}

Before carrying out any numerical computations, 
let us analyze the asymptotic behaviour of \eqref{eq:RH} as 
$\nu \to 0^+$, $\nu \to \infty$, $\alpha\to 1^+$, and $\alpha\to\infty$.  
The results, shown in Table~\ref{tab:limits}, 
are valid for all parametrizations of \eqref{eq:reduced} and 
do not rely on \eqref{eq:parametrization}. 

\begin{table}[H]
    \tbl{Limits of $\Ieq_+$, $\Req_+$, and the sign of $y_\nu(\alpha)$.}
    \centering
    {\begin{tabular}{@{}lrrr@{}} \toprule
         $\lim$ \qquad \qquad \qquad ~ 
         &\qquad ~ $\Ieq_+$
         &\qquad ~ $\Req_+$
         &\qquad ~ $y_\nu(\alpha)$\\ 
         \midrule
         $\nu \to 0^+$ 
         & $\frac{4 c_1 c_2 + c_3}{4 \beta c_2(\gamma + \mu)}$ 
         & $\frac{\gamma(\beta - (\gamma + \mu))(\mu + \omega \kappa)}{\beta c_2}$ 
         & $> 0$ \\
         \midrule
         $\nu \to \infty$ 
         & $\frac{c_1}{\beta(\gamma+\mu)}$ 
         & $\frac{c_1\gamma}{\beta\mu(\gamma+\mu)}$ 
         & $> 0$ \\
         \midrule
         $\alpha \to 1^+$ 
         & $\frac{(\kappa+\mu)(\beta-(\gamma+\mu))}{\beta(\gamma+\mu+\kappa)}$ 
         & $\frac{\gamma(\beta-(\gamma+\mu))}{\beta(\gamma+\mu+\kappa)}$ 
         & $> 0$ \\
         \midrule
         $\alpha \to \infty$ 
         & $\frac{(\kappa+\mu)(\beta-(\gamma+\mu))}{\beta(\gamma+\mu+\kappa)}$
         & $0$ 
         & $> 0$ \\
         \bottomrule
    \end{tabular}}
    \begin{tablenotes}
      \small
      \item \textsuperscript{*} The details of the computations are to be 
        found in Appendix~\ref{app:asymptotic}.
    \end{tablenotes}
    \label{tab:limits}
\end{table}

As a consequence of these limits, 
there exists a compact region $K$ in the $(\alpha, \nu)$-plane 
such that the endemic equilibrium $\ee$ is LAS for 
$(\alpha, \nu) \in (1, \infty) \times (0, \infty) \setminus K$. 

\subsubsection{Double bubbles of instability}
\label{sec:RH-EE-numerics}
\begin{figure}[!ht]
\centering
\begin{subfigure}{.54\textwidth}
    \centering
    \includegraphics[scale=0.78]{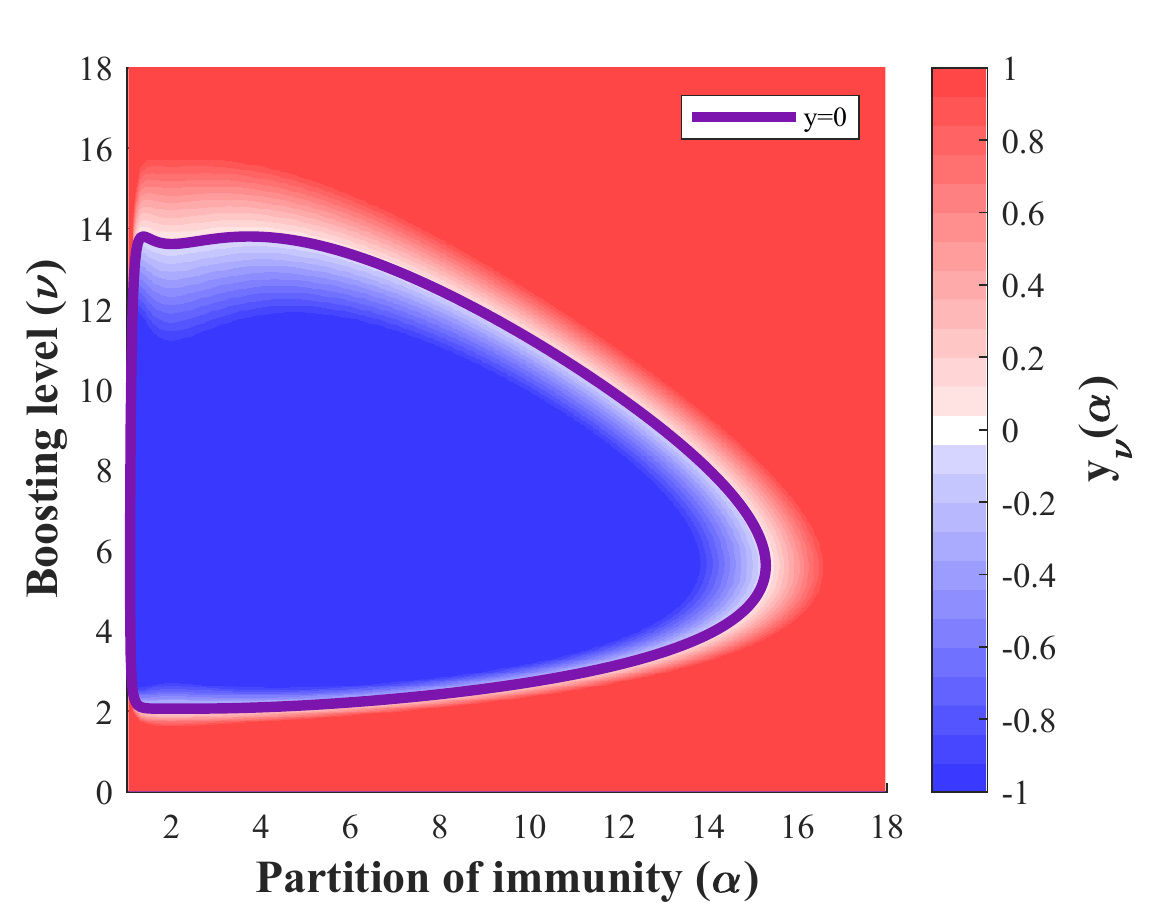}
    \captionsetup{
        margin = 20pt
    }
    \caption{}
    \label{fig:RH-Large-Left}
\end{subfigure}~\qquad
\begin{subfigure}{.4\textwidth}
    \centering
    \includegraphics[scale=0.78]{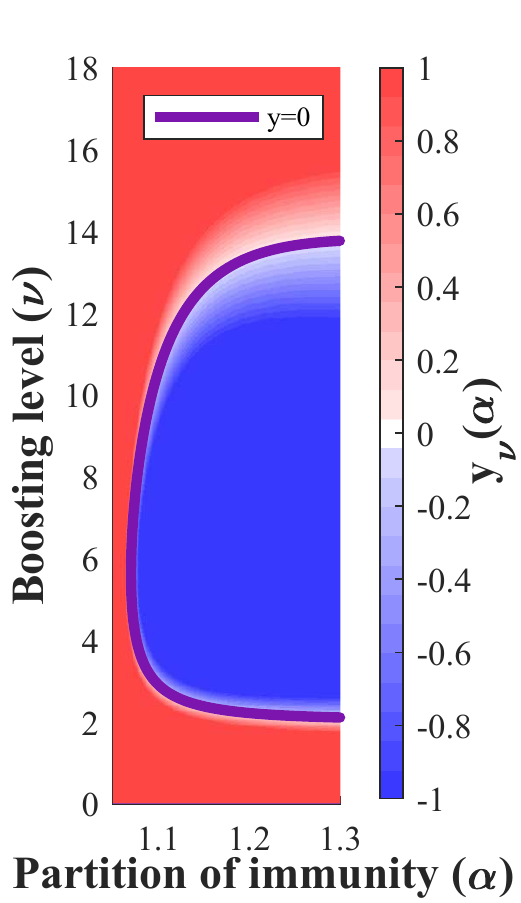}
    \captionsetup{
        margin = 20pt
    }
    \caption{}
    \label{fig:RH-Large-Right}
\end{subfigure}
    \captionsetup{
        margin = 0pt
    }
    \caption{Heatmap of the Routh-Hurwitz criterion $y_\nu(\alpha)$ capped at $[-1, 1]$ with 
	highlighted zero contour. Figure~\ref{fig:RH-Large-Right} zooms in on the region close to $\alpha = 1$.}
    \label{fig:RH-Large}
\end{figure}

Section~\ref{sec:LAS-EE-numerics} has readily established  that 
it is sufficient to consider a compact subset in the $(\alpha, \nu)$-plane 
for the stability analysis of $\ee$. Based on our experiments, we have 
restricted our attention to $(\alpha, \nu) \in [1.01, 18] \times [0.01, 18]$ 
and obtained the heatmap in Figure~\ref{fig:RH-Large} when studying the 
positivity of $y_\nu(\alpha)$. 

It is apparent that, for an interval of $\alpha$ values, 
$y_\nu(\alpha)$ is initially positive for small $\nu$, then, as the boosting rate increases the RH criterion becomes negative for an interval of $\nu$ values, 
after which, it turns positive again.
Now, let us look at the heatmap from the other direction. 
Note that for most interesting boosting rates $\nu$, 
a similar stability switch may be observed over an $\alpha$-interval. 
However, the dynamics is clearly more involved close to 
boosting rates around $14$ as 
Figure~\ref{fig:RH-Large} suggests the presence of multiple stability switches. 

\begin{figure}[H]
\centering
\begin{subfigure}{.54\textwidth}
    \centering
    \includegraphics[scale=0.78]{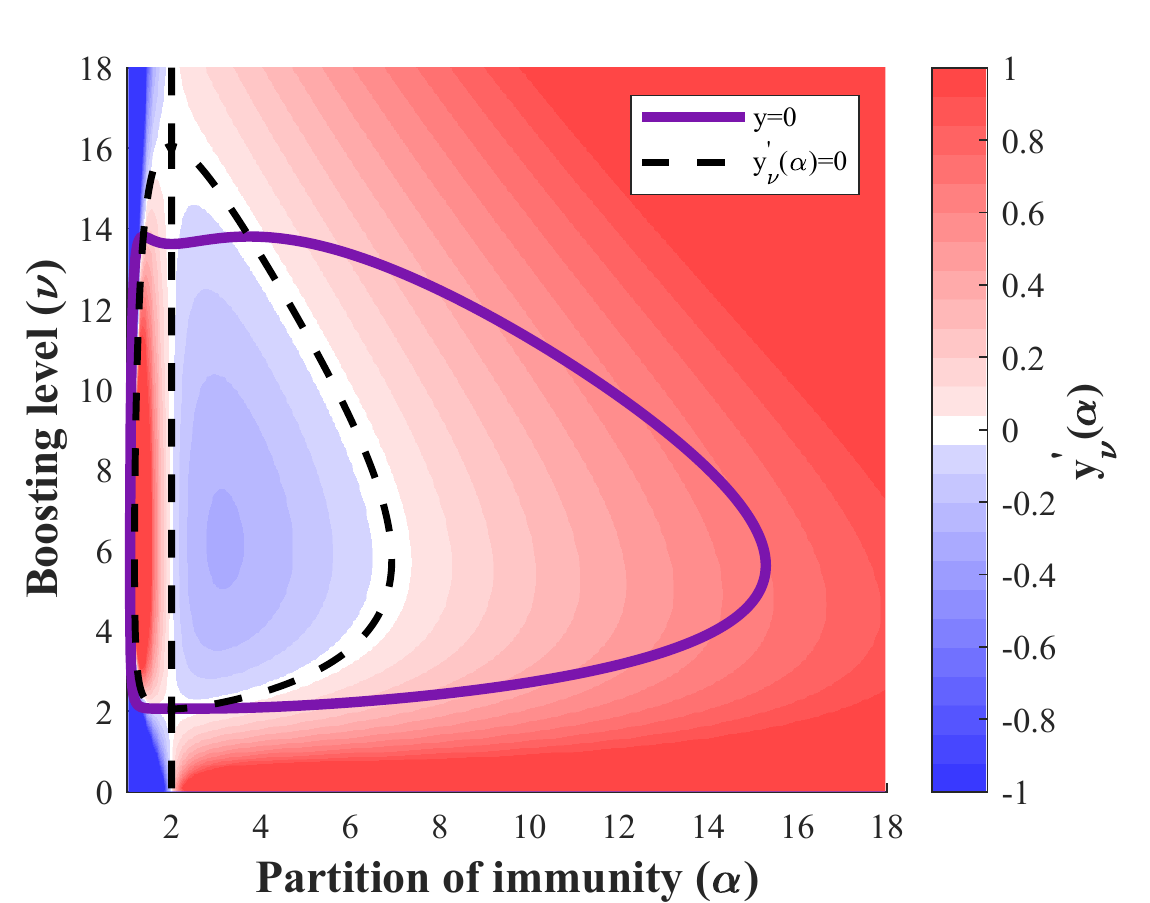}
    \captionsetup{
        margin = 20pt
    }
    \caption{}
    \label{fig:RH-Deriv-Left}
\end{subfigure}\qquad
\begin{subfigure}{.4\textwidth}
    \centering
    \includegraphics[scale=0.78]{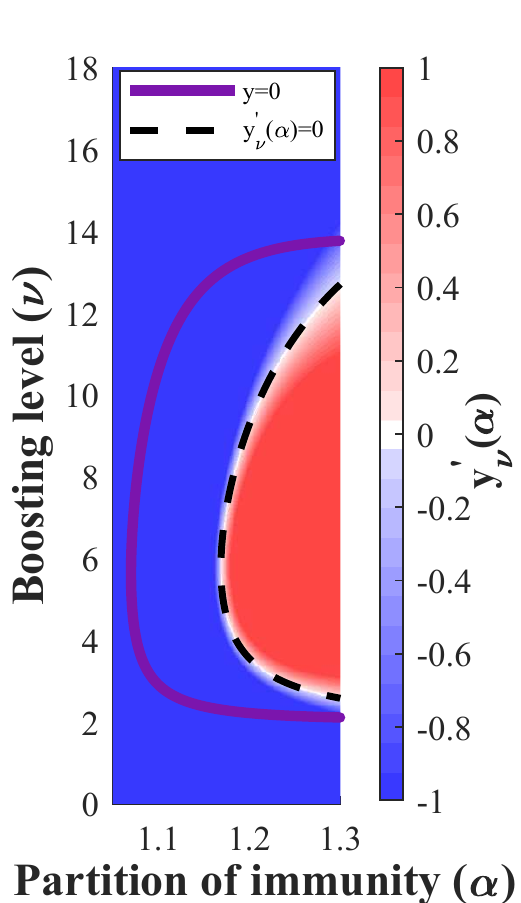}
    \captionsetup{
        margin = 0pt
    }
    \caption{}
    \label{fig:RH-Deriv-Right}
\end{subfigure}
    \captionsetup{
        margin = 0pt
    }
    \caption{Heatmap of $y'_\nu(\alpha)$ capped at $[-1, 1]$ with 
	highlighted zero contours of $y_\nu(\alpha)$ and $y'_\nu(\alpha)$.  Figure~\ref{fig:RH-Deriv-Right} zooms in on the region close to $\alpha = 1$.}
    \label{fig:RH-Deriv}
\end{figure}

It is straightforward to localize such phenomena 
by finding local extrema of $y_\nu(\alpha)$ (as a function of $\alpha$) whose 
value is zero. Hence, we looked for 
intersections of the curves
$$y_\nu(\alpha) = 0 \qquad \qquad \mbox{and} \qquad \qquad 
    \frac{\partial}{\partial \alpha} y_\nu(\alpha) \equiv y'_\nu(\alpha) = 0$$
as shown in Figure~\ref{fig:RH-Deriv} together with the positivity 
analysis of the derivative. Our findings
confirm the presence of multiple switches 
close to $\nu \approx 13.7$, moreover, they highlight  
the existence of similar dynamics close to 
$\nu \approx 2.06362$ as well. 
Note that Figure~\ref{fig:RH-Large} gives no hint of the latter. 

\begin{figure}[ht]
\centering
\begin{subfigure}{0.47\textwidth}
    \centering
    \includegraphics[scale=0.8]{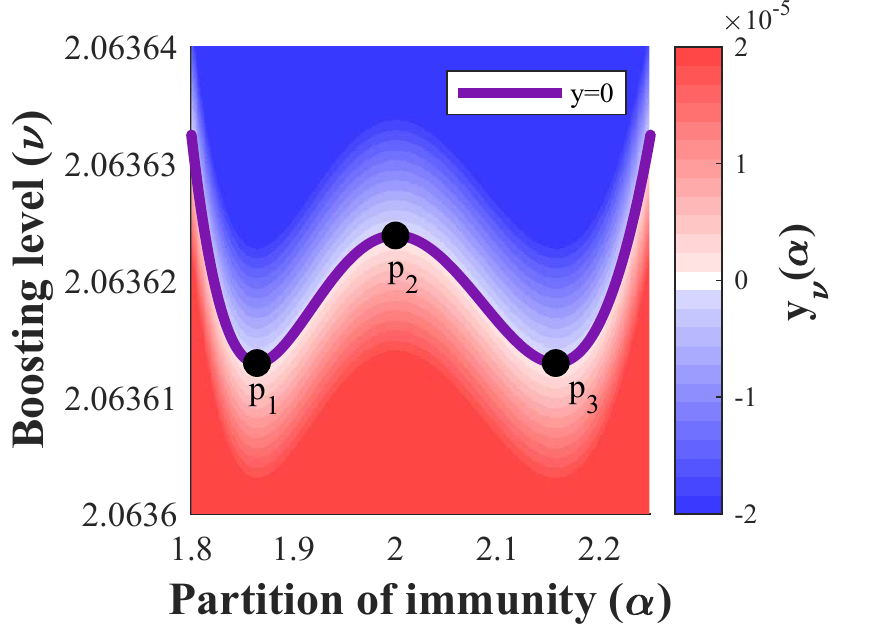}
    \captionsetup{
        margin = 20pt
    }
    \caption{$\nu \approx 2.06362$.}
    \label{fig:RH-bubbles-nu-2_06}
\end{subfigure}\qquad
\begin{subfigure}{.47\textwidth}
    \centering
    \includegraphics[scale=0.78]{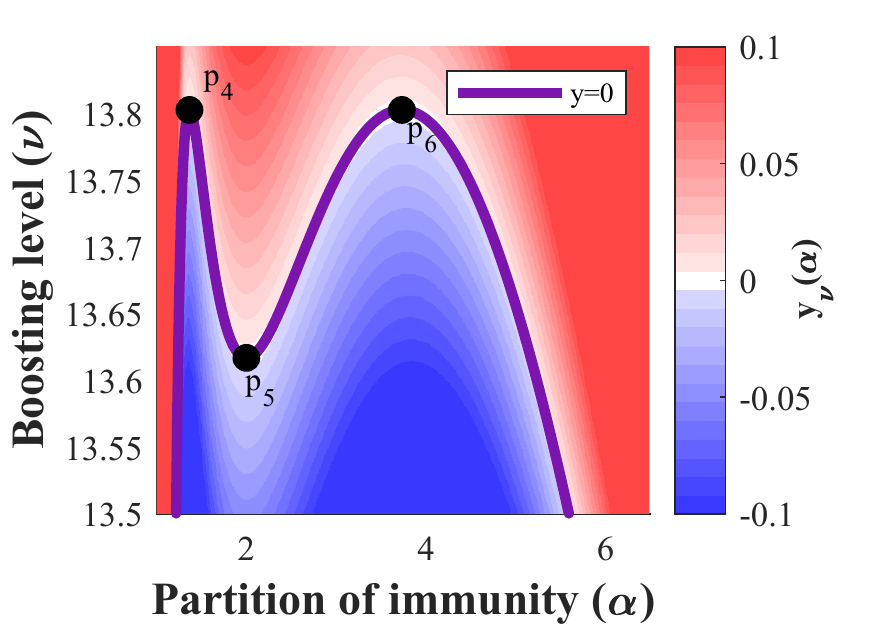}
    \captionsetup{
        margin = 20pt
    }
    \caption{$\nu \approx 13.7$.}
    \label{fig:RH-bubbles-nu-13_7}
\end{subfigure}
    \captionsetup{
        margin = 0pt
    }
    \caption{Zoomed-in heatmaps of the Routh-Hurwitz criterion 
	$y_\nu(\alpha)$ with 
	highlighted zero contour over regions of interest in the 
	$(\alpha, \nu)$-plane. Critical points 
	$p_i = (\alpha^*_i, \nu^*_i)$ 
	on the contour are marked. }
    \label{fig:RH-bubbles}
\end{figure}

Recall from Section~\ref{sec:LAS-EE-Eta} that local extrema of $y_\nu(\alpha)$ 
-- other than $\alpha = 2$ -- appear in pairs. 
Hence, zooming in on these 
two regions, shown in Figure~\ref{fig:RH-bubbles}, reveals double bubbles  
of instability for certain boosting rates.

Note that the width of the $\nu$-range where this phenomenon occurs 
in Figure~\ref{fig:RH-bubbles-nu-2_06} is less than $2 \cdot 10^{-5}$, thus, it 
should come as no surprise that it was not observable 
based on the original heatmap in Figure~\ref{fig:RH-Large}. 
The coordinates of the highlighted critical points are given in
Table~\ref{app:tab-critical-points}. 

\subsection{Numerical bifurcation analysis}
\label{sec:bif-alpha}

In the following, we present numerical analysis  
of one parameter $(\alpha)$ and 
two parameter $(\alpha, \nu)$ bifurcations 
of the endemic equilibria branch 
carried out 
using \matcont \cite{MatCont_article}. For a background on bifurcation analysis we refer to \cite{kuznetsov2013elements, wiggins2003introduction}.

\begin{figure}[H]
	\centering
		\includegraphics[scale=.8]{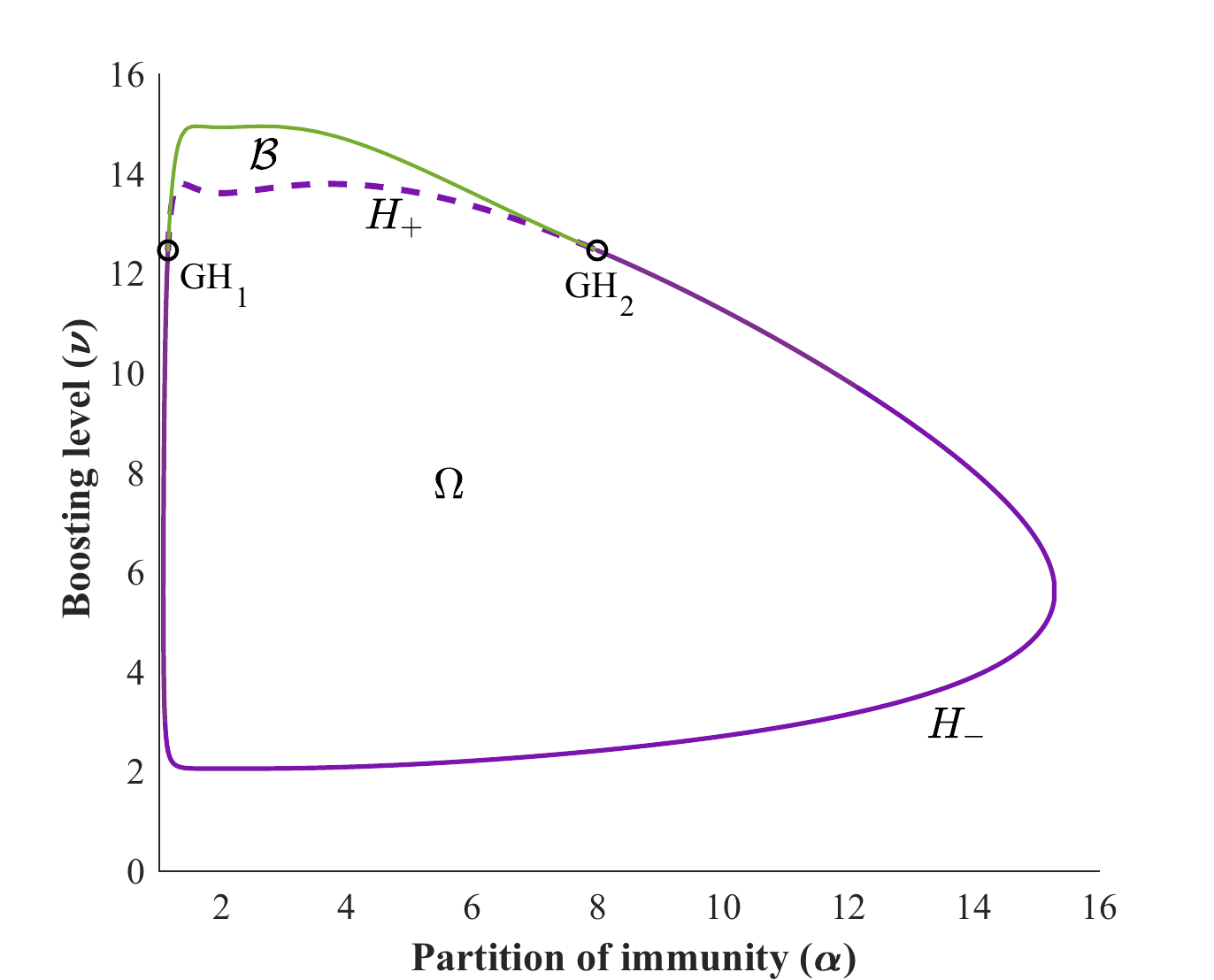}
	\caption{Two-parameter $(\alpha,\nu)$ bifurcation diagram.}\label{fig:bif-diagram-large-alpha-nu}
\end{figure}

Motivated by the results of Section~\ref{sec:LAS-EE-numerics},
in particular the region 
depicted in Figure~\ref{fig:RH-Large-Left}, 
we computed the two parameter $(\alpha,\nu)$ bifurcation 
diagram of system \eqref{eq:reduced}, see 
Figure~\ref{fig:bif-diagram-large-alpha-nu}. To fully understand the bifurcation diagram, let us denote by $\Omega$ the open domain enclosed by the purple colored Hopf curve, which is continuous when supercritical (called $H_{-}$) and dashed when subcritical (called $H_{+}$). A stable limit
cycle bifurcates from the equilibrium if we cross $H_{-}$ from outside to inside $\Omega$,
while an unstable cycle appears if we cross $H_{+}$ in the opposite direction.

It is apparent that for larger boosting rates 
($\nu$ between $12$ -- $15$), 
the local stability analysis of $\ee$ is not sufficient to capture 
all interesting dynamics.

The two new critical points identified 
are $\gh_1 = (\alpha^*_{\gh_1}, \nu^*_{\gh_1})$ and 
$\gh_2 = (\alpha^*_{\gh_2}, \nu^*_{\gh_2} \equiv \nu^*_{\gh_1})$. The approximate 
coordinates of these \emph{generalized Hopf points} 
are listed in 
Table~\ref{app:tab-critical-gh-points} and they mark the parameter values where the Hopf bifurcation
changes from supercritical to subcritical. The branch of the limit points of periodic cycles appears in green, which together with the dashed purple curve $H_{+}$ enclose a bistability region $\mathcal{B}$, where there exists a stable periodic solution 
alongside the LAS endemic equilibrium.

Let us now examine the bifurcation diagram in more detail  
over regions,
characterized by various levels of boosting rate $\nu$,
where the dynamics is similar. 

In all bifurcation plots that follow, the endemic equilibria branch (particularly the $I$ component) is marked with black curve, solid when LAS and dashed when unstable. Red and blue curves represent branches of stable and unstable limit cycles, respectively, and Hopf bifurcation points are marked with purple dots.

\paragraph*{Region: $0 \leq \nu < \nu_1^* \equiv \nu_3^*$.}
The system has a stable point attractor 
for all $\alpha > 1$.

\begin{figure}[H]
	\centering
	\includegraphics[scale=.8]{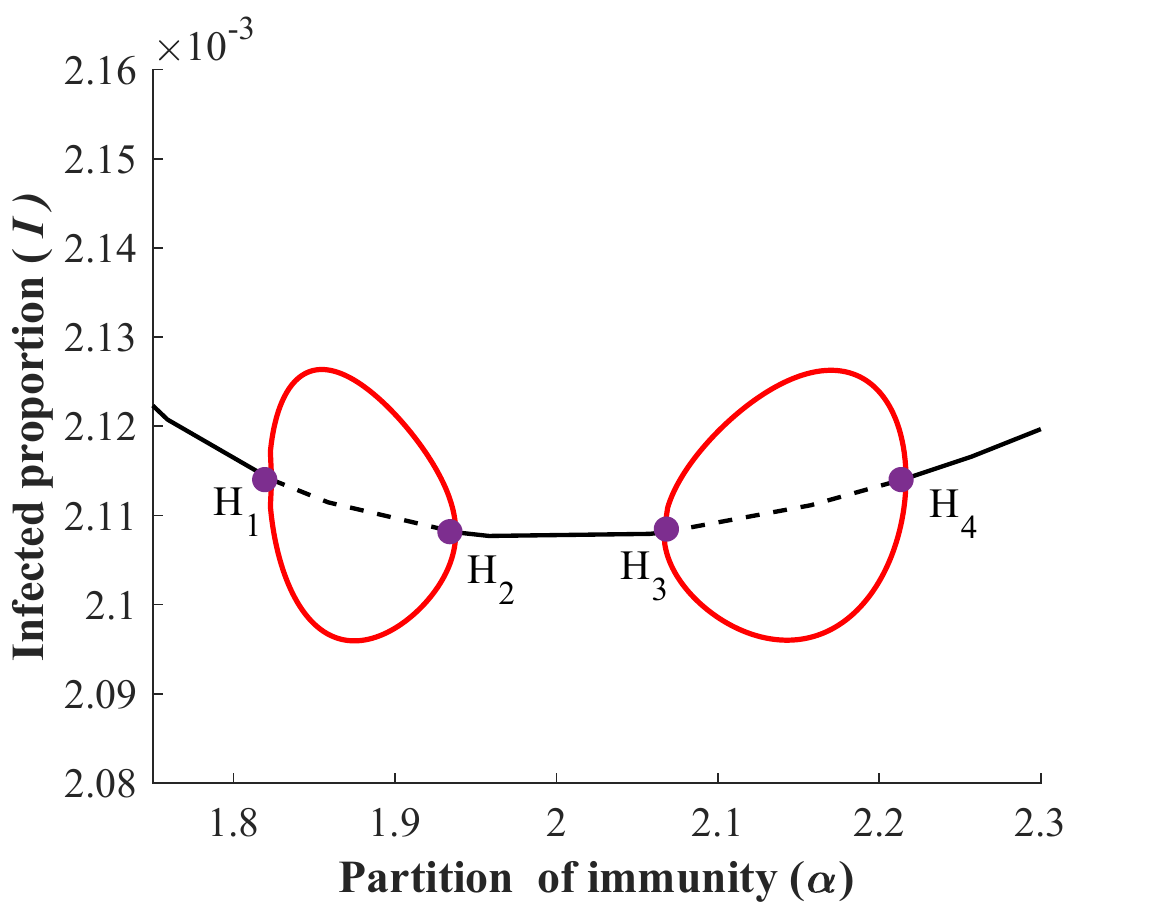}
	\label{fig:bif-diagram-alpha-nu-2_06}
    \caption{Bifurcation diagram w.r.t. $\alpha$, 
    when $\nu = 2.06362$. }
    \label{bif-diagram-bubbles-nu-2_06}
\end{figure}

\paragraph*{Region: $\nu_1^* \equiv \nu_3^* < \nu < \nu_2^*$.}
There are four supercritical Hopf bifurcation points on the endemic equilibria branch, 
see Figure~\ref{bif-diagram-bubbles-nu-2_06} 
for a typical setting. Continuation of (the $I$-component of) limit cycles with respect to $\alpha$ starting
from two Hopf bifurcation points, $\h_1$ and $\h_2$, forms an \emph{endemic bubble} (the two branches of stable limit cycles coincide), see \cite{bubble} for the origin of this concept. The same happens for the $\h_3, \h_4$ pair.

Recall that these double bubbles of instability 
(endemic bubbles) were 
readily observed in Figure~\ref{fig:RH-bubbles-nu-2_06}. Such double bubbles have been conjectured in a delay differential model for waning and boosting \cite{doublebubble}. For an overview of similar phenomena, the 
reader is referred to 
\cite{leblanc2016degenerate, sherborne2018bursting, krisztin2011bubbles}. 

\begin{figure}[H]
\centering
\begin{subfigure}{.47\textwidth}
    \centering
    \includegraphics[scale=0.75]{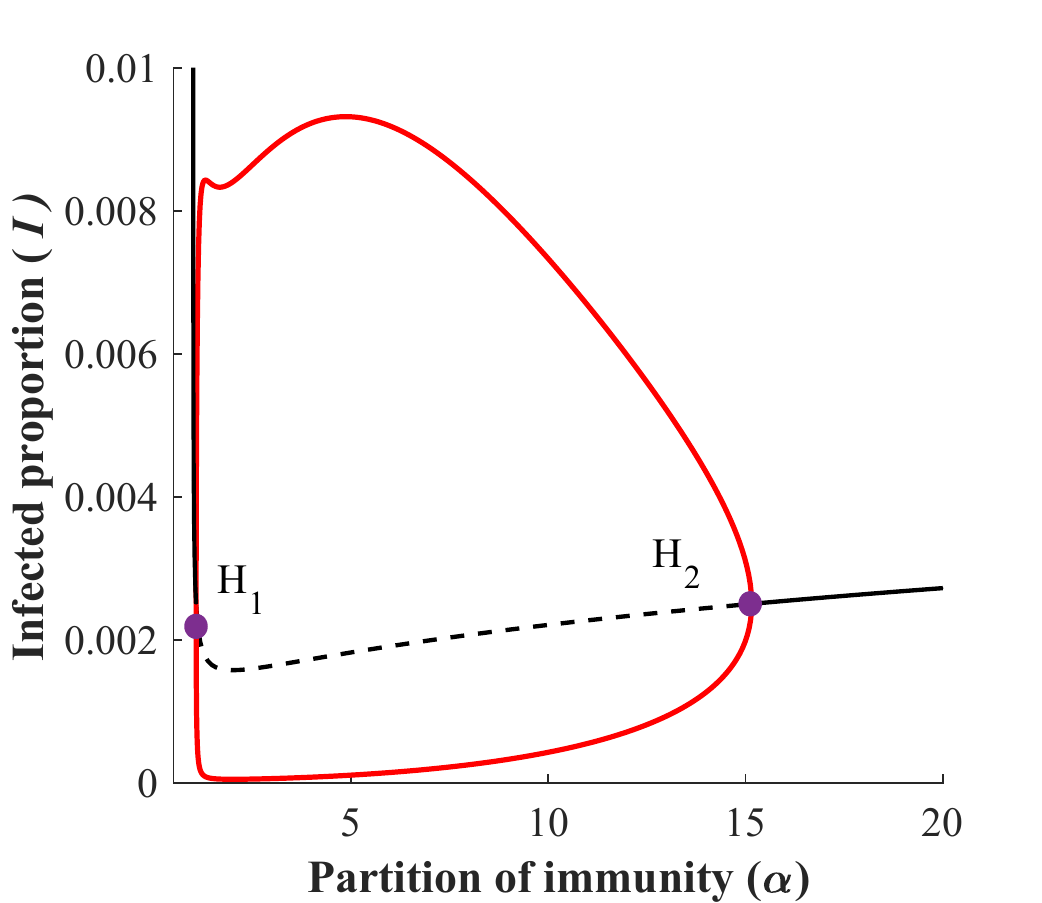}
    \caption{}
    \label{fig:bif-diagram-bubble-nu-5_8-large}
\end{subfigure}\qquad
\begin{subfigure}{.45\textwidth}
    \centering
    \includegraphics[scale=0.75]{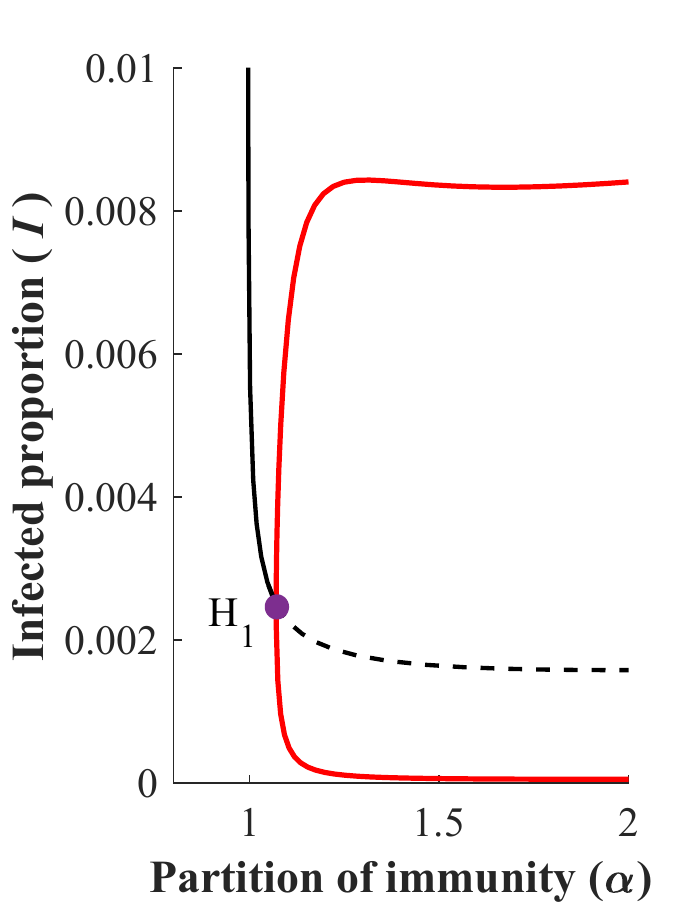}
    \captionsetup{
        margin = 0pt
    }
    \caption{}
    \label{fig:bif-diagram-bubble-nu-5_8-zoomed}
\end{subfigure}
    \captionsetup{
        margin = 0pt
    }
    \caption{(a) Bifurcation diagram w.r.t. $\alpha$,
    when $\nu = 5.8$; (b) Zoom of (a) close to the vertical line $\alpha = 1$.}
    \label{fig:bif-diagram-bubble-nu-5_8}
\end{figure}

\paragraph*{Region: $\nu^*_2 < \nu < \nu^*_{\gh_1} 
\equiv \nu^*_{\gh_2}$.}
As the boosting rate increases, 
the middle supercritical Hopf points $\h_2$ and $\h_3$ (observed in the 
previous region) get closer to each other, finally collide 
and we obtain a single endemic bubble, Figure~\ref{fig:bif-diagram-bubble-nu-5_8}. 

\paragraph*{Region: $\nu^*_{\gh_1} \equiv \nu^*_{\gh_2} < 
\nu < \nu^*_5$.}

As $\nu$ continues to grow in the two-parameter plane  in Figure~\ref{fig:bif-diagram-large-alpha-nu}, two generalized Hopf points, $\gh_1$ and $\gh_2$, appear. They separate branches of sub- and supercritical Hopf bifurcations in the parameter plain. The stable limit cycles survive when we enter region $\mathcal{B}$. Crossing the subcritical Hopf boundary $H_+$ creates an extra unstable cycle inside the first one, while the equilibrium regains its stability. Two cycles of opposite stability exist inside the bistable region $\mathcal{B}$ and disappear at the green curve. 

\begin{figure}[H]
	\centering
	\begin{subfigure}{0.47\textwidth}
		\includegraphics[scale=0.75]{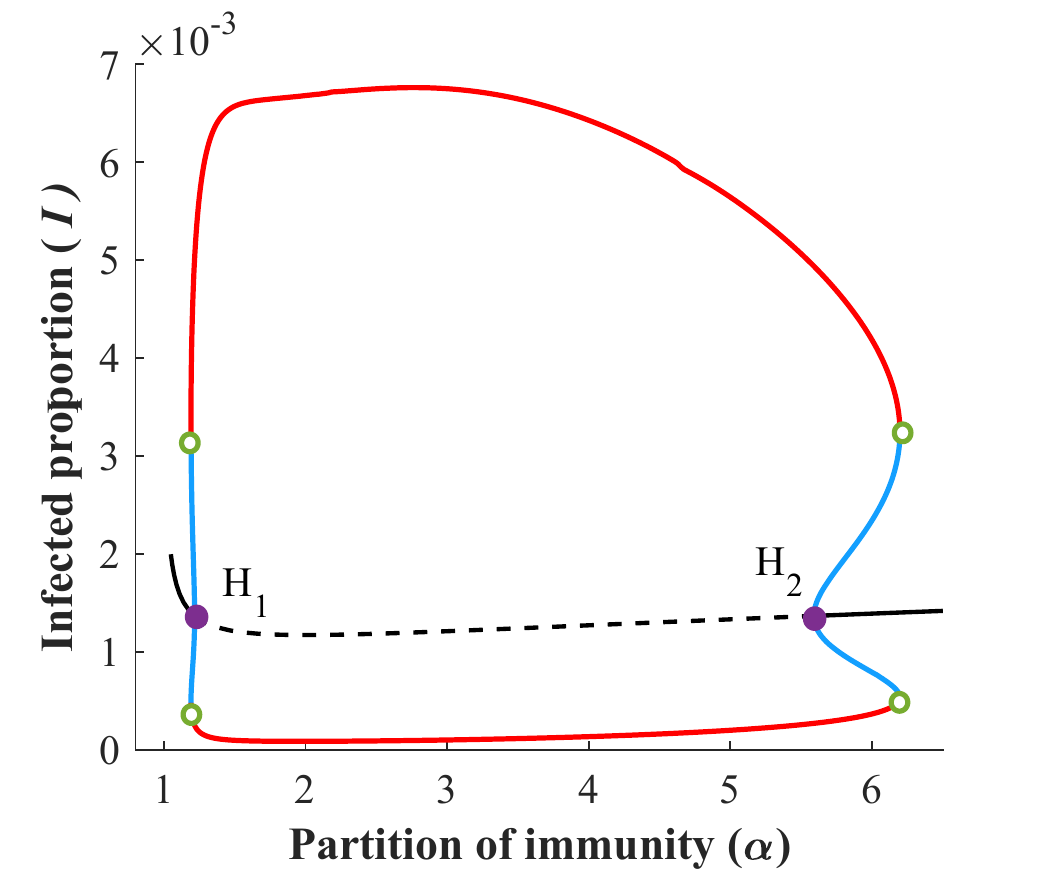}
	\end{subfigure}\qquad
	\begin{subfigure}{0.45\textwidth}
		\includegraphics[scale=0.75]{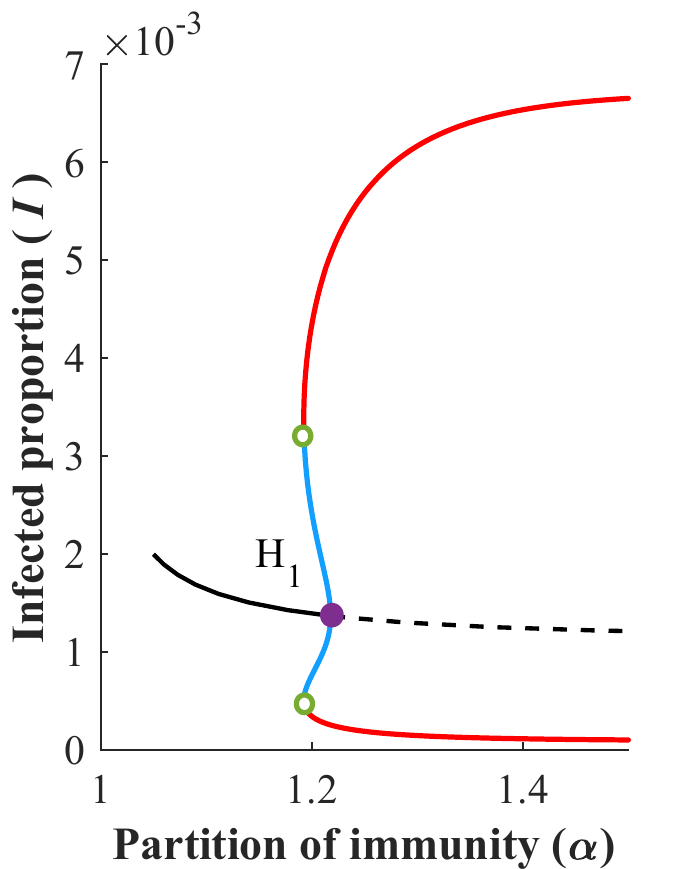}
	\end{subfigure}
	\caption{Bifurcation diagram w.r.t $\alpha$, with $\nu=13.5$ (left) and zoom into the bistable region around $H_1$ (right).} \label{fig:bif_nu=13_5new}
\end{figure}

When we pass the generalized Hopf points and fix a $\nu$ in this region, then Figure~\ref{fig:bif_nu=13_5new} shows a typical bifurcation w.r.t. $\alpha$. Observe here the two small $\alpha$-parameter ranges of bistability where the EE and the larger amplitude periodic solution are both stable. The points marked with green circle are limit points of periodic orbits. The stable and unstable cycles collide and disappear on the green curve in Figure~\ref{fig:bif-diagram-large-alpha-nu}, corresponding to a fold bifurcation of cycles. 

\paragraph*{Region: $\nu^*_5 < \nu < \nu^*_4 \equiv \nu^*_6$.}

\begin{figure}[H]
	\centering
	\begin{minipage}{0.48\textwidth}
		\includegraphics[width=200pt]{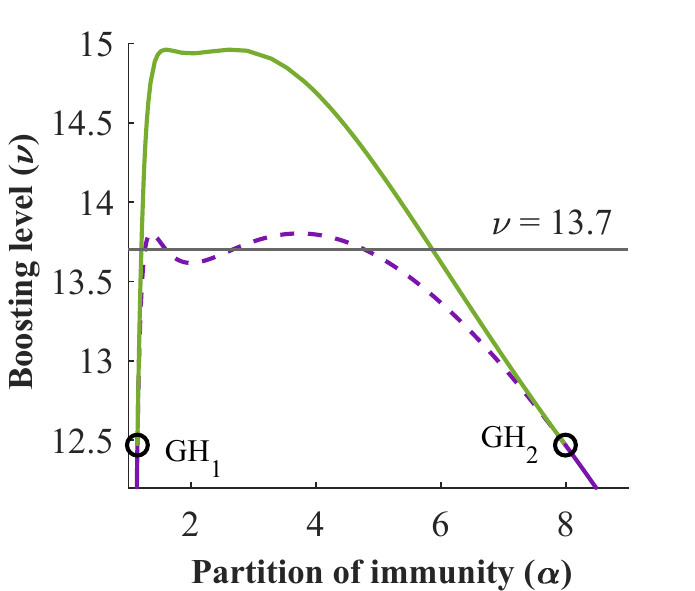}
	\end{minipage}
	\begin{minipage}{0.48\textwidth}
		\includegraphics[width=200pt]{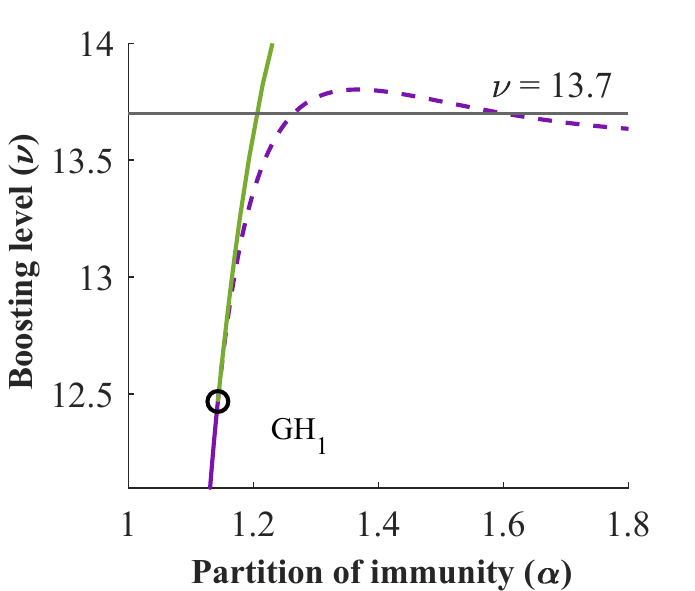}
	\end{minipage}
	\caption{Two-parameter $(\alpha,\nu)$ bifurcation diagram, bistability region.}\label{fig:bistability_zoom}
\end{figure}

As we increase the boosting value, the dynamics is changing, as observed on the shape of the subcritical Hopf curve $H_{+}$ in Figure~\ref{fig:bistability_zoom} and the heat map in Figure~\ref{fig:RH-bubbles-nu-13_7}. 

\begin{figure}[H]
	\centering
	\begin{subfigure}{0.47\textwidth}
		\includegraphics[scale=0.75]{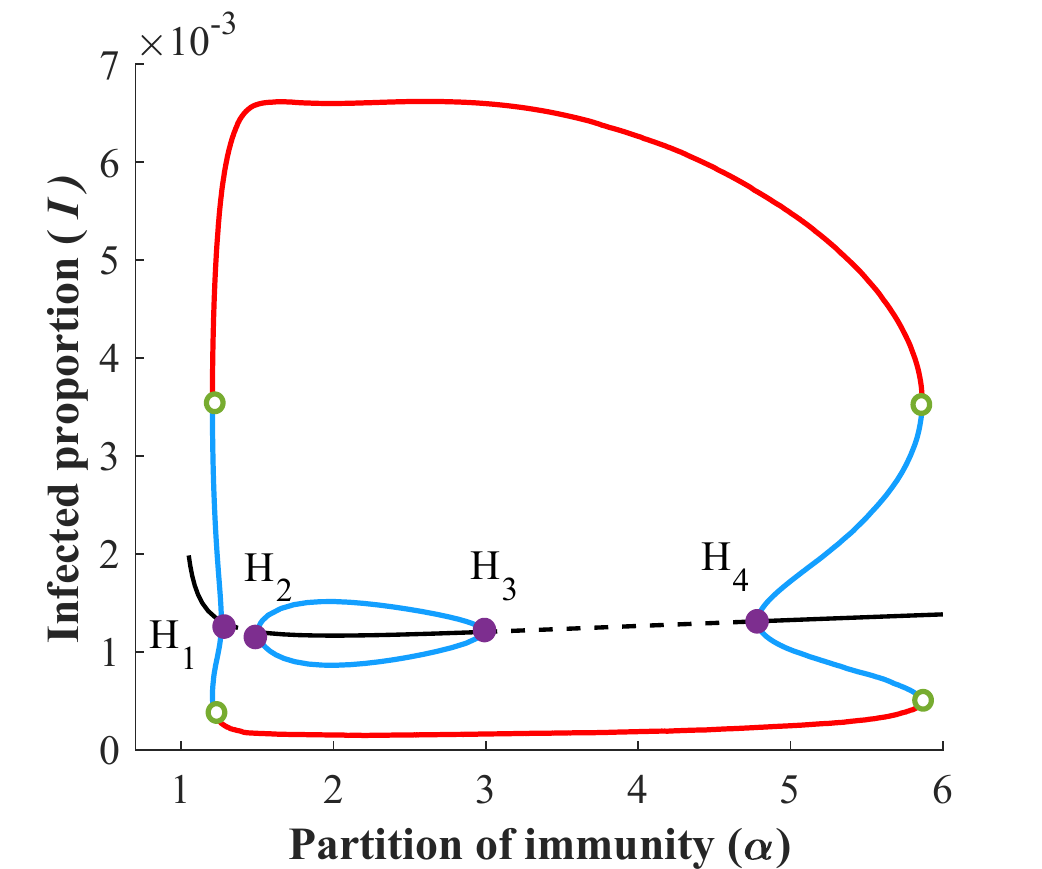}
	\end{subfigure}\qquad
	\begin{subfigure}{0.45\textwidth}
		\includegraphics[scale=0.75]{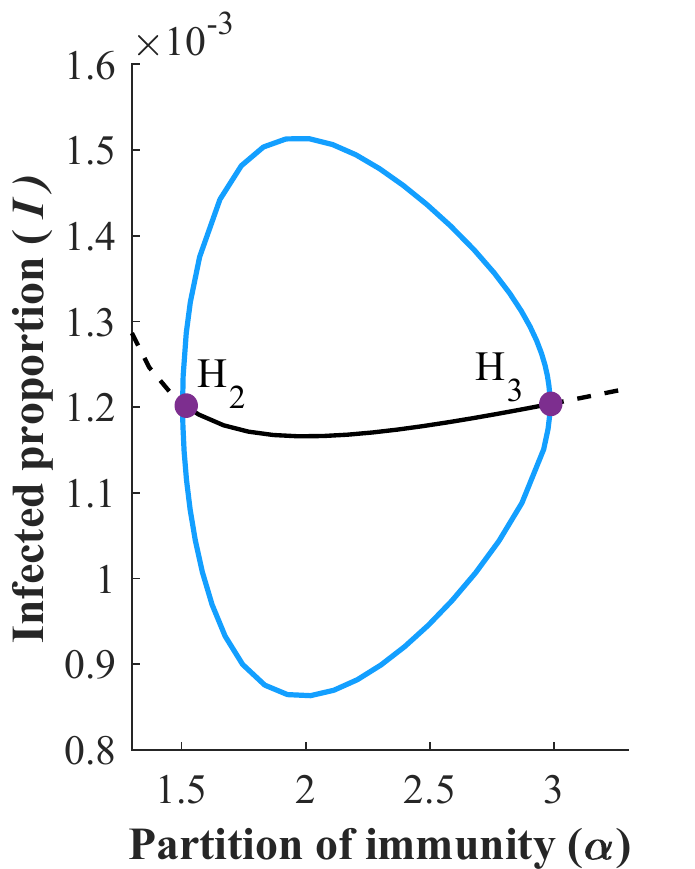}
	\end{subfigure}
	\caption{Bifurcation diagram w.r.t $\alpha$, with $\nu=13.7$ (left) and zoomed into the bubble (right).} \label{fig:bif_nu=13.7}
\end{figure}

In Figure~\ref{fig:bif_nu=13.7}, the bifurcation diagram confirms the existence of four subcritical Hopf bifurcation points. Here a small bubble appears inside the region of stable oscillations, which leads to an additional bistable region compared to the previous case.  

When we increase the boosting in this region, i.e., still intersecting the subcritical Hopf curve, the Hopf points $\h_1$ and $\h_2$ as well as $\h_3$ and $\h_4$ move closer to each other, resulting in larger bistability regions, see also  the heatmap Figure~\ref{fig:RH-bubbles-nu-13_7}.   

\begin{figure}[H]
	\centering
	\includegraphics[scale=0.8]{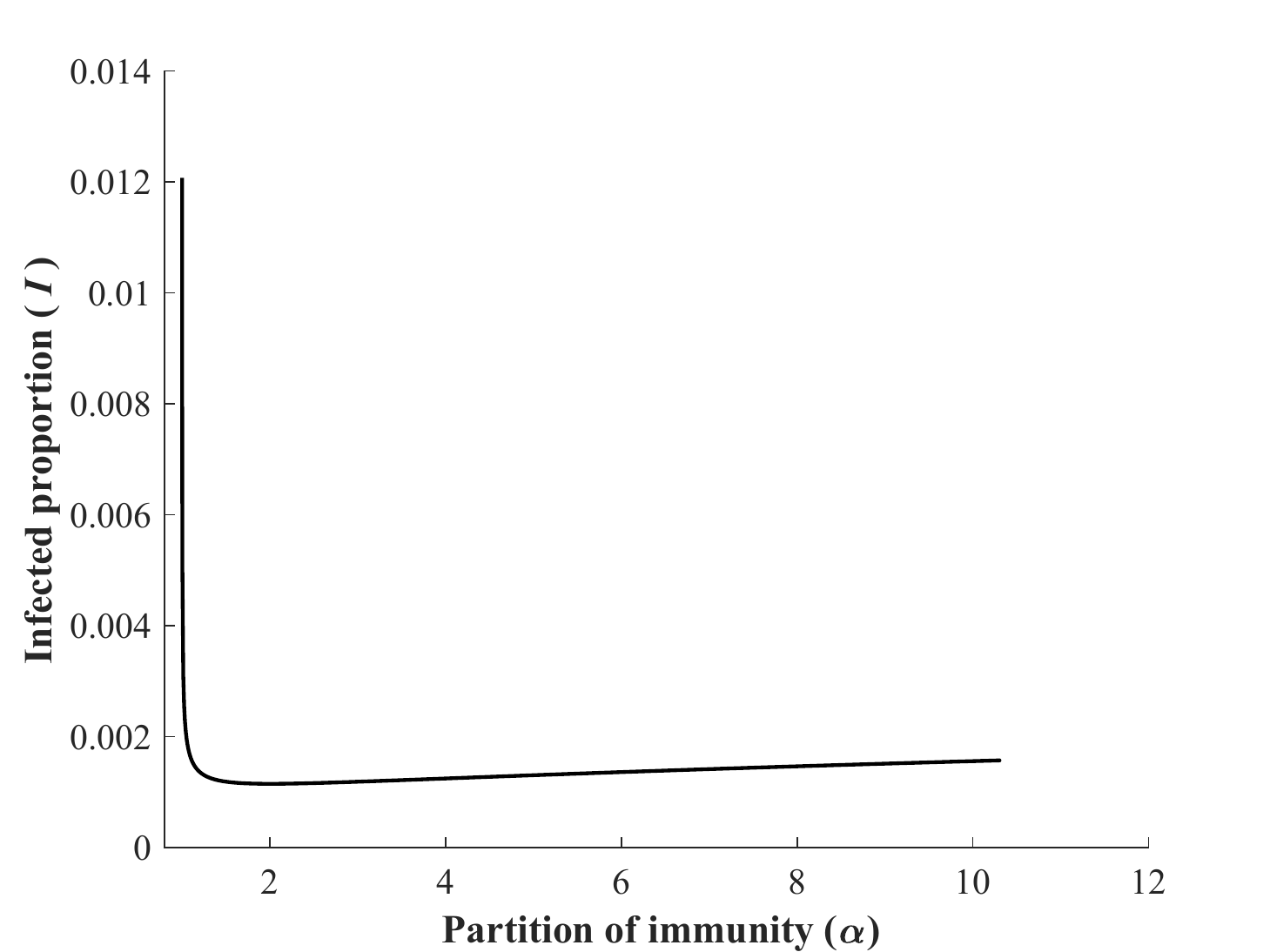}
	\caption{There are no bifurcations of equilibria when $\nu=14.5$.}\label{fig:bif_nu=14.5}
\end{figure}

\paragraph*{Region: $\nu^*_4 \equiv \nu^*_6 < \nu$.}
As we enter this region we leave $H_{+}$ and do not intersect any Hopf branches, hence, the continuation method 
utilized so far leaves us with a single
stable equilibrium, Figure~\ref{fig:bif_nu=14.5}.

\begin{figure}[H]
	\centering
	\includegraphics[scale=0.8]{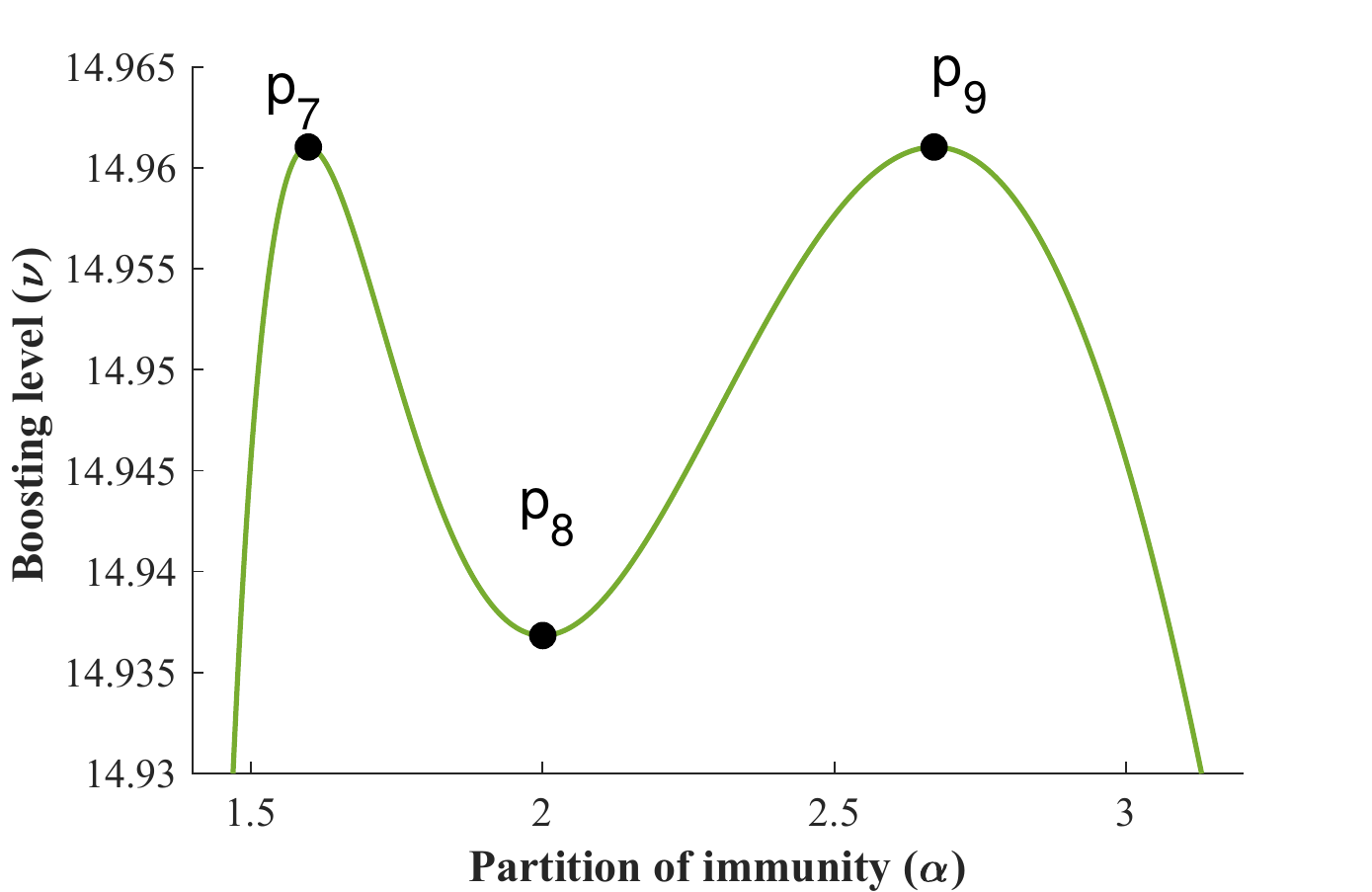}
	\caption{Branch of the limit points of periodic solutions.  }\label{fig:greencurve}
\end{figure}

There is however, a range of $\nu$ values in this region that belong to $\mathcal{B}$,  as observed in Figure~\ref{fig:bif-diagram-large-alpha-nu}. For a better demonstration of the shape of the limit cycle branch, see Figure~\ref{fig:greencurve}. The coordinates of the critical points can be found in Table~\ref{app:tab-critical-points}.

\begin{figure}[H]
	\centering
	\includegraphics[scale=0.8]{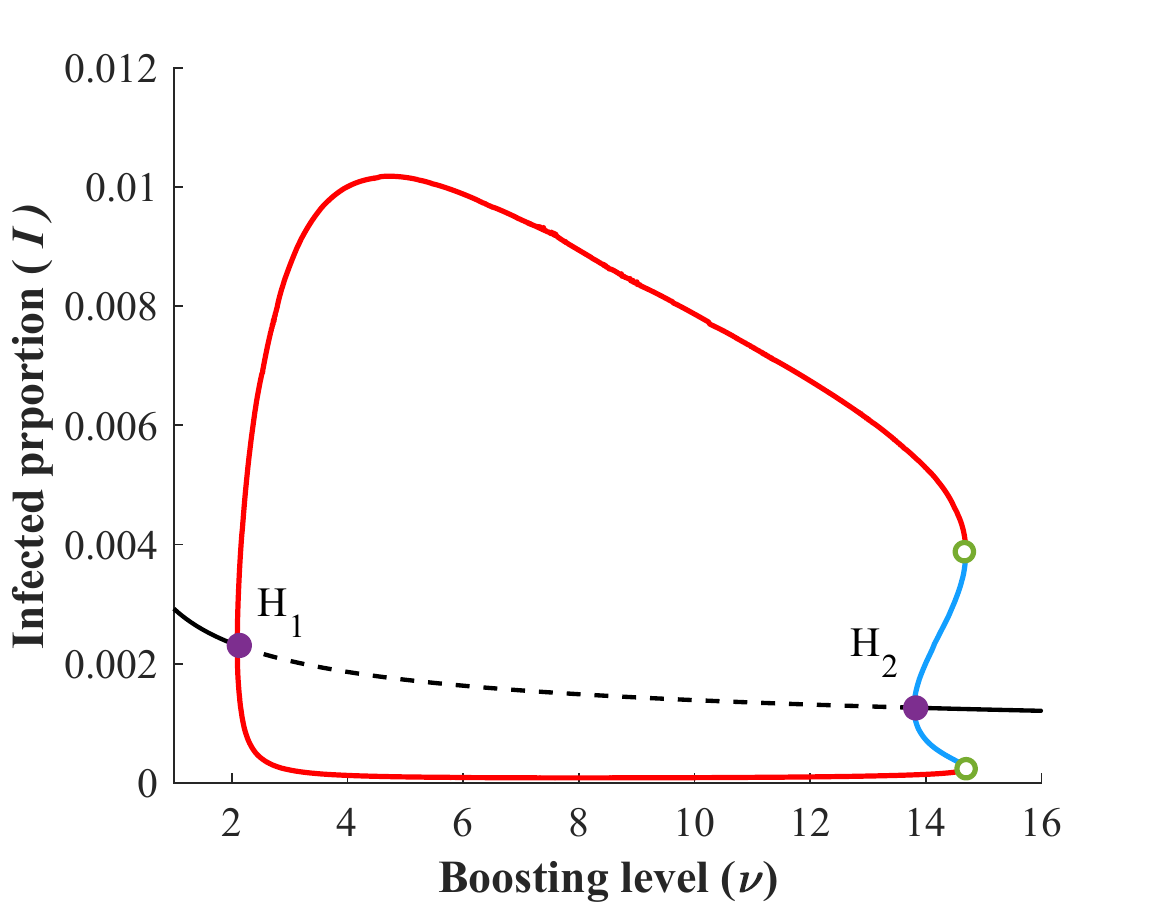}
	\caption{Bifurcation diagram w.r.t. $\nu$,
    when $\alpha=4$.}\label{fig:bif_alpha=4}
\end{figure}

Considering the heatmaps in Figure~\ref{fig:RH-bubbles}, it was natural to investigate regions in the two parameter plane $(\alpha,\nu)$ where $\nu$ is constant and look at bifurcations with respect to $\alpha$. To capture the extension of the bistability region in the $\nu$ direction we can investigate 
the dynamics for $\alpha$ fixed and consider the boosting rate 
$\nu$ as the bifurcation parameter. For a typical setting see Figure~\ref{fig:bif_alpha=4}.  

\vspace{2em}

\section{Conclusions}
We generalized previous compartmental SIRWS models of waning and boosting of immunity by allowing different expected durations for individuals being in the fully immune compartment $R$ and being in the waning immunity compartment $W$, from where their immunity can still be restored  upon re-exposure. 
We proposed an asymmetric division of the immunity period in the SIRWS model to these two phases, characterized by a newly introduced bifurcation parameter. Other parameters were chosen to mimic pertussis. 
We observed and established a new symmetry in 
these divisions around 
the critical case of equal partitioning when analyzing the 
stability criterion of the endemic equilibrium. 
This, combined with numerical bifurcation methods, enabled us to 
characterize the model dynamics for a 
relevant range of parameter values.  We composed global bifurcation diagrams, and found complex and rich dynamics where stability switches, Hopf bifurcations, folds of periodic branches appeared, forming interesting structures in the parameter space. We found double endemic bubbles as well as regions of bistability. This study confirmed that simple looking SIRWS ODE models can have very intricate dynamics. Our analyis highlighted that the division of the immunity period into maximally immune and boostable phases is a key parameter, which significantly determines the dynamics of the system. As a consequence, future epidemiological studies should attempt to estimate this quantity to have a better description of the influence of waning-boosting mechanisms on epidemic outcomes.

\section*{Acknowledgement}

This research was supported by the Ministry of Innovation and
Technology of Hungary from the National Research, Development
and Innovation Fund, project no. TKP2021-NVA-09. 
In addition, F.B. and G.R. were supported by NKFIH (FK 138924, KKP 129877).
M.P. was also supported by the Hungarian Scientific Research Fund, Grant No. K129322 and SNN125119; 
F.B. was also supported by UNKP-21-5 and the 
Bolyai Scholarship of the 
Hungarian Academy of Sciences.


\vspace{2em}

\appendix

\section*{Appendices}
\addcontentsline{toc}{section}{Appendices}
\label{sec:appendices}

\renewcommand*{\thesubsection}{\textsc{\textbf{A\arabic{subsection}}}}

\renewcommand{\thetheorem}{A\arabic{subsection}.\arabic{theorem}}

\renewcommand{\thelemma}{A\arabic{subsection}.\arabic{lemma}}

\renewcommand{\thetable}{A\arabic{subsection}.\arabic{table}}

\renewcommand{\theequation}{A\arabic{subsection}.\arabic{equation}}

\renewcommand{\thefigure}{A\arabic{subsection}.\arabic{figure}}

\setcounter{lemma}{0}
\setcounter{table}{0}
\setcounter{equation}{0}
\setcounter{figure}{0}
\setcounter{theorem}{0}

\subsection{Derivation of the formula for 
\texorpdfstring{{\boldmath$\Req_{\pm}$}}{R*}}
\label{app:Rstar}

We now derive the formula for $\Req_{\pm}$ used in 
Section \ref{sec:EE-analytic}.
As we have seen, substituting the 
formulae for $\Seq$ and $\Ieq$ into \eqref{eq:equilib_2}, 
we obtain the quadratic equation 
\begin{equation*}
A (R^*)^2 + B R^* + C = 0, 
\end{equation*}
with coefficients 
\begin{align*}
A &= \frac{
        \omega \kappa \nu \beta (\gamma + \mu)
    }{
        (\gamma + \mu + \omega \kappa)^2
    }, \\
B &= \frac{
        \omega \kappa \nu (\gamma + \mu + c_0 c_1) - [
            (\gamma + \mu) 
            (\mu + \alpha \kappa + \omega \kappa + \nu c_0 c_1) + 
            \omega \kappa (\alpha \kappa + \beta \nu)
        ]
    }{
        \gamma + \mu + \omega \kappa
    }, \\
C &= \frac{c_0 c_1}{\beta}
        [\beta \nu + \gamma - \gamma \nu - \mu \nu - c_0 c_1 \nu].
\end{align*}
The y-intercept C may be simplified as 
\begin{align*}
C &= \frac{c_0 c_1}{\beta} 
    \left(
        \beta \nu + \gamma - \gamma \nu - \mu \nu - c_0 c_1 \nu 
    \right), \\
  &= c_0 c_1 \nu + \frac{c_0 c_1 \gamma}{\beta} - 
    \frac{c_0 c_1 \nu}{\beta}
    \left(
        \gamma + \mu + c_0 c_1
    \right), \\
  &= c_0 c_1 \nu + \frac{c_0 c_1 \gamma}{\beta} - 
  \frac{c_0 c_1 \nu}{\beta} 
  (
    \beta - \frac{c_0 c_1 \gamma}{\omega \kappa + \mu} - 
    c_0 c_1 + c_0 c_1
  ), \\
  &= \frac{c_0 c_1 \gamma}{\beta} + 
    \frac{c_0^2 c_1^2 \gamma \nu}{\beta (\omega \kappa + \mu)}, \\
  &= \frac{c_0 c_1 \gamma}{\beta} 
    \left(
        1 + \frac{c_0 c_1 \nu}{\omega \kappa + \mu}
    \right).
\end{align*}
Then, the solution formula gives
\begin{equation}
\label{eq:app-Rstar-quadratic-formula}
\begin{split}
    R_{\pm}^{*} = \frac{-B}{2A} \mp \frac{\sqrt{B^2 - 4 A C}}{2A}. 
\end{split}
\end{equation}
We split \eqref{eq:app-Rstar-quadratic-formula} 
into two parts and evaluate them separately. 
\begin{align*}
&\frac{-B}{2A} 
    = \frac{
            \gamma + \mu + \omega \kappa
        }{
            2 \omega \kappa \nu \beta (\gamma + \mu)
        } 
        \left( 
            (\gamma + \mu) 
            (\mu + \alpha \kappa + \omega \kappa + \nu c_0 c_1) + 
            \omega \kappa (\alpha \kappa + \beta \nu) - 
            \omega \kappa \nu (\gamma + \mu + c_0 c_1) 
        \right) \\
    &= \frac{
            \gamma + \mu + \omega \kappa
        }{
            2 \omega \kappa \nu \beta (\gamma + \mu)
        }
        \left( 
            (\gamma + \mu) (\alpha \kappa + \omega \kappa + \mu) + 
            (\gamma + \mu) \nu c_0 c_1 + \omega \alpha \kappa^2 +  
            \omega \kappa \beta \nu - \omega \kappa \nu 
            \left(
                \beta - \frac{\gamma c_0 c_1}{\omega \kappa + \mu}
            \right) 
        \right) \\
    &= \frac{ 
            \gamma + \mu + \omega \kappa
        }{
            2 \omega \kappa \beta
        }
        \left(
            \frac{
                (\gamma + \mu)
                (\alpha \kappa + \omega \kappa + \mu) + 
                \omega \alpha \kappa^2
            }{
                \nu (\gamma + \mu)
            } + \frac{
                (\gamma + \mu) \nu c_0 c_1
            }{
                \nu (\gamma + \mu)
            } + \frac{
                \omega \kappa \nu \gamma c_0 c_1 
            }{
                \nu (\gamma + \mu) (\omega \kappa + \mu)
            }
        \right) \\
    &= \frac{
            \gamma + \mu + \omega \kappa
        }{
            2 \omega \kappa \beta
        }
        \left(
            \frac{c_2}{\nu (\gamma + \mu)} + 
            c_0 c_1 + 
            \frac{
                \omega \kappa \gamma c_0 c_1 
            }{
                (\omega \kappa + \mu)
                (\gamma + \mu)
            }
        \right) \\
    &= \frac{
            \gamma + \mu + \omega \kappa
        }{
            2 \omega \kappa \beta
        }
        \left(
            \frac{c_2}{\nu (\gamma + \mu)} +  c_0 c_1 + 
            \frac{
                \omega \kappa \gamma (\beta - \gamma - \mu) 
            }{
                (\omega \kappa + \mu + \gamma)
                (\gamma +\mu)
            }
        \right) \\
    &= \frac{
            \gamma + \mu + \omega \kappa 
        }{
            2 \omega \kappa \beta
        }
        \left(
            \frac{c_2}{\nu (\gamma + \mu)} + 
            c_0 c_1 + c_0 c_1 - \frac{c_1}{\gamma + \mu}
        \right) \\
    &= \frac{
            \gamma + \mu + \omega \kappa
        }{
            2 \omega \kappa \beta
        }
        \left(
            \frac{c_2}{\nu (\gamma + \mu)} +  2 c_0 c_1 - 
            \frac{c_1}{\gamma + \mu}
        \right).
\end{align*}
Then, the other term in \eqref{eq:app-Rstar-quadratic-formula} is 
\begin{align*}
\mp & \frac{ \sqrt{B^2 - 4 A C} }{2 A} = \\
    &= \mp \frac{
            (\gamma + \mu + \omega \kappa)^2
        }{
            2 \omega \kappa \nu \beta  (\gamma + \mu)
        }
        \sqrt{
            \frac{
                (\mu (\beta - \gamma - \mu) )^2 
            }{
                (\gamma + \mu + \omega \kappa)^2
            } 
            \nu^2 + 
            \frac{T_0}{(\gamma + \mu + \omega \kappa)^2} \nu + 
            \frac{ 
                ((\gamma + \mu) (\alpha \kappa + \omega \kappa+\mu) + 
                \tiny \alpha \omega \kappa^2)^2
            }{
                (\gamma + \mu + \omega \kappa)^2
            }
        }, \\
\end{align*}
with 
\begin{align*}
    T_0 &= 2 (\beta - ( \gamma + \mu) )( 
        \gamma \mu^2 + \mu^3 + \omega \kappa \mu^2 + 
        \alpha \kappa \mu^2 + \alpha \gamma \kappa \mu +  
        \omega \gamma \kappa \mu + 2 \alpha \omega \gamma \kappa^2 +  
        \alpha \omega \mu \kappa^2
    ) \\
    &= 4 \gamma \alpha \omega \kappa^2 (\beta - (\gamma + \mu)) + 
    2 \mu (\beta - (\gamma + \mu)) 
    [
        (\gamma + \mu) 
        (\omega \kappa + \alpha \kappa + \mu) + 
        \alpha \omega \kappa^2
    ] \\
    &= c_3 + 2 c_1 c_2.
\end{align*}
Hence, 
\begin{align*}
\mp & \frac{ \sqrt{B^2 - 4 A C} }{2 A} = \\
    &= \mp \frac{
            (\gamma + \mu + \omega \kappa)^2
        }{
            2 \omega \kappa \nu \beta (\gamma + \mu)
        }
        \sqrt{
            \frac{c_1^2}{(\gamma + \mu + \omega \kappa)^2} \nu^2 + 
            \frac{
                (2 c_1 c_2 + c_3)
            }{
                (\gamma + \mu + \omega \kappa)^2
            } \nu + 
            \frac{c_2^2}{(\gamma + \mu + \omega \kappa)^2}
        } \\
    &= \mp \frac{
            \gamma + \mu + \omega \kappa
        }{
            2 \omega \kappa \nu \beta (\gamma + \mu)
        }
        \sqrt{c_1^2 \nu^2 + c_2^2 + 2 \nu c_1 c_2 + c_3 \nu} \\
    &= \mp \frac{
            \gamma + \mu + \omega \kappa
        }{
            2 \omega \kappa \nu \beta (\gamma + \mu)
        }
        \sqrt{(c_1 \nu +c_2)^2 + c_3 \nu}.
\end{align*}
Recombining the two terms yields the formula  
\begin{equation*}
\begin{split}
    R_{\pm}^{*} 
    &= \frac{
        \gamma + \mu + \omega \kappa
    }{
        2 \beta \omega \kappa
    } 
    \left[ 
        \left( 
            2 c_0 - \frac{1}{\gamma + \mu}
        \right) c_1 + 
        \frac{1}{\nu (\gamma + \mu)} \left(
            c_2 \mp \sqrt{(c_1 \nu + c_2)^2 + c_3 \nu}
        \right) 
    \right].
\end{split}
\end{equation*}

\setcounter{lemma}{0}
\setcounter{table}{0}
\setcounter{equation}{0}
\setcounter{figure}{0}
\setcounter{theorem}{0}

\subsection{Transcritical bifurcation of forward type}\label{app:transcritical}

For the sake of completeness, we include a slightly adjusted 
version of Theorem 4.1. from Castillo-Chavez and Song 
\cite{castillo2004dynamical}.

\begin{theorem}
\label{app:thm-transcritical}
Let $f\in C^{2}(\mathbb{R}^n\times \mathbb{R},\mathbb{R}^n)$ and 
consider the system of ordinary differential equations 
\begin{equation*}
    \frac{dx}{dt} = f(\xvec, b),
\end{equation*}
with $b$ as a parameter.
Assume that $\mathbf{0}$ is an equilibrium point, 
i.e., $f(\mathbf{0}, b) = 0$ for all $b \in \mathbb{R}$. 
In addition, assume the following:

\begin{itemize}
    \item[(i)] The linearization of the system at $(\mathbf{0}, 0)$ 
    \begin{equation*}
        A := D_{\xvec} f(\mathbf{0}, 0) = \left(
            \frac{\partial f_{i}}{\partial \xvec_{j}}(\mathbf{0}, 0)
        \right)_{i, j = 1}^n    
    \end{equation*}
    has zero as a simple eigenvalue and all other eigenvalues of $A$ have negative real parts.
    
    \item[(ii)] The matrix $A$ has a non-negative right eigenvector $w$ and a left eigenvector $v$ corresponding to the zero eigenvalue.
\end{itemize}

Let $f_{k}$ be the $k$-th component of $f$ and define 
\begin{align*}
    Z_1 &= \sum_{k, i, j = 1}^{n} v_{k} w_{i} w_{j} 
    \frac{
        \partial^{2} f_{k}
    }{
        \partial \xvec_{i} \partial \xvec_{j}
    }(\mathbf{0}, 0) 
    \qquad \mbox{and} \\ 
    Z_2 & =\sum_{k, i = 1}^{n} v_{k} w_{i} 
    \frac{
        \partial^{2} f_{k}
    }{
        \partial \xvec_{i} \partial b
    }(\mathbf{0}, 0). 
\end{align*}

If $Z_1 < 0$ and $Z_2 > 0$, then 
as $b$ changes from negative to positive, 
the equilibrium $\mathbf{0}$ changes its stability 
from stable to unstable. 
At the same time, a negative unstable equilibrium 
becomes positive and locally asymptotically stable. 
Hence, a forward bifurcation occurs at $b = 0$.
\end{theorem}

\setcounter{lemma}{0}
\setcounter{table}{0}
\setcounter{equation}{0}
\setcounter{figure}{0}
\setcounter{theorem}{0}

\subsection{Transformation of \texorpdfstring{$y_{\nu}(\alpha)$}{the Routh-Hurwitz criterion}}
\label{app:RH-EE-Eta}

The alternative, simpler form of 
$\omega \kappa \beta \nu (1 - \Seq - \Ieq_+ - \Req_+)$,
used in Section \ref{sec:LAS-EE-Eta}, is obtained as follows.
\begin{align*}
    \omega \kappa &\beta \nu (1 - \Seq - \Ieq_+ - \Req_+) = \\
    &~ \omega \kappa \beta \nu \left(1 - \frac{\gamma + \mu}{\beta} \right) - 
    \omega \kappa \beta \nu \frac{\sqrt{(c_1 \nu + c_2)^2 + c_3 \nu} + 
        (c_1 \nu - c_2)}{2 \beta \nu (\gamma + \mu)} \\
    ~ &- \omega \kappa \beta \nu \frac{\gamma + \mu + \omega \kappa}{2 \beta \omega \kappa} \left[ 
    \left( 
        2 c_0 - 
        \frac{1}{\gamma + \mu}
    \right) c_1 + 
    \frac{1}{
        \nu (\gamma + \mu)}
    \left(
        c_2 - \sqrt{(c_1 \nu + c_2)^2 + c_3 \nu}
    \right) \right] \\
    = \, &\omega \kappa \nu \frac{c_1}{\mu} + \frac{\omega \kappa}{2} 
    \frac{c_2 - \sqrt{(c_1 \nu + c_2)^2 + c_3 \nu}}{\gamma + \mu} - 
        \frac{\omega \kappa}{2} 
    \frac{c_1 \nu}{\gamma + \mu} - 
        \frac{\nu}{2} (\gamma + \mu + \omega \kappa) 2 c_0 c1 \\
    & - \frac{\nu (\gamma + \mu)}{2} \left[ 
         - \frac{1}{\gamma + \mu} c_1 + 
    \frac{1}{
        \nu (\gamma + \mu)}
    \left(
        c_2 - \sqrt{(c_1 \nu + c_2)^2 + c_3 \nu}
    \right) \right] \\
    &+ \frac{\nu \cdot \omega \kappa}{2} \frac{c_1}{\gamma + \mu} - 
    \frac{\omega \kappa}{2} 
    \frac{c_2 - \sqrt{(c_1 \nu + c_2)^2 + c_3 \nu}}{\gamma + \mu} \\
    = \, & \omega \kappa \nu \frac{c_1}{\mu} - 
        \frac{\nu}{2} (\gamma + \mu + \omega \kappa) 2 c_0 c1 
        + \frac{\nu}{2} c_1 - \frac{1}{2} \left(
        c_2 - \sqrt{(c_1 \nu + c_2)^2 + c_3 \nu}
        \right) \\
    = \, & \omega \kappa \nu \frac{c_1}{\mu} - 
        \nu \left(1 + \frac{\omega \kappa}{\mu} \right) c1 
        + \frac{\nu}{2} c_1 - \frac{1}{2} \left(
        c_2 - \sqrt{(c_1 \nu + c_2)^2 + c_3 \nu}
        \right) \\ 
    = \, & \frac{ \sqrt{(c_1 \nu + c_2)^2 + c_3 \nu} - (c_1 \nu + c_2) }{2} 
    = \beta \nu (\gamma + \mu) \Ieq_+ - c_1 \nu. \\
\end{align*}

We now present two Lemmas on derivatives of function compositions. 
The first is a version of the classical result by Faà di Bruno 
generalizing the chain rule.

\begin{lemma}[Faà di Bruno]
\label{app:lemma-faa}
Let $f \colon I \to U$ and $g \colon U \to V$ be analytic functions, 
where $I, U, V \subseteq R$ are connected subsets. 
Consider the Taylor expansions 
$f(t) = \sum_{k = 0}^\infty \taylor{f}{k} (t - t_0)^k$ 
centered at $t_0 \in I$ with $t \in I$ and 
$g(x) = \sum_{k = 0}^\infty \taylor{g}{k} (x - x_0)^k$ 
centered at $x_0 = f(t_0)$ for $x \in U$. Then, 
the composite function $(g \circ f)$ attains the Taylor expansion 
$(g \circ f)(t) = \sum_{k = 0}^\infty \taylor{g \circ f}{k} (t - t_0)^k$ 
centered at $t_0$ with the coefficients 
\begin{equation*}
\begin{split}
    \taylor{g \circ f}{0} &= \taylor{g}{0}
        \qquad \mbox{and} \\
    \taylor{g \circ f}{k} &= 
        \sum_{\substack{
            b_1 + 2 b_2 + \ldots + k b_k = k \\
            m := b_1 + b_2 + \ldots + b_k
        }} ~ 
        \frac{m!}{b_1! b_2! \ldots b_k!} \taylor{g}{m} 
        \prod_{i = 1}^k 
        \Big( \taylor{f}{i} \Big)^{b_i},
\end{split}
\end{equation*}
where $k \geq 1$ and 
$b_1, \ldots, b_k$ are nonnegative integers. 
\end{lemma}

Using the results of Lemma~\ref{app:lemma-faa} and assuming 
that the inner function has a vanishing first derivative and 
the outer function 
has a cascade of vanishing derivatives, the following Lemma 
establishes a similar 
property for the composite function. 

\begin{lemma}
\label{app:lemma-composite-zeros}
Assume that $f$ and $g$ are as in Lemma~\ref{app:lemma-faa} and that 
$\taylor{f}{1} = 0$. Then,
\begin{enumerate}[label=(\alph*)]
    \item $\taylor{g \circ f}{2} = 0$ $\Longleftrightarrow$ 
    $\taylor{g}{1} = 0$,
    \item if $\taylor{g}{i} = 0$ for $i = 1, \ldots, k - 1$, then 
    $\taylor{g \circ f}{2k} = 0$ 
    $\Longleftrightarrow$ $\taylor{g}{k} = 0$,
    \item if $\taylor{g}{i} = 0$ for $i = 1, \ldots, k$, then 
    $\taylor{g \circ f}{2k + 1} = 0$.
\end{enumerate}
\end{lemma}

\begin{proof}
The claims directly follow from Lemma~\ref{app:lemma-faa} by noting 
that in the formula of $\taylor{g \circ f}{k}$, for terms with 
$m > k / 2$, the 
inequality $b_1 > 0$ must hold, hence, any such term must evaluate to zero.
\end{proof}

\setcounter{lemma}{0}
\setcounter{table}{0}
\setcounter{equation}{0}
\setcounter{figure}{0}
\setcounter{theorem}{0}

\subsection{Asymptotic behaviour of equilibria}
\label{app:asymptotic}
The analytic computations of the behavior of the equilibria of SIRWS system for large and small boosting ($\nu$) 

    \[  \lim_{\nu \to 0^+} \Ieq_+= 
    \frac{(\alpha\kappa + \mu)(\gamma+\mu+\omega\kappa )c_1c_0}{\beta c_2 },\quad  
     \lim_{\nu\to\infty}\Ieq_+
    =\frac{\abs{c_1}+c_1}{2\beta(\gamma+\mu)},
\]

\[
\begin{split}
    \lim_{\nu\to 0}\Req_+ 
    &= \frac{\gamma + \mu + \omega \kappa}{2 \beta \omega \kappa} \left[ 
    \left( 
        2 c_0 - 
        \frac{1}{\gamma + \mu}
    \right) c_1 -
    \frac{2c_1c_2+c_3}{
         2c_2(\gamma + \mu)}
     \right],
\end{split}
\]

\[
\begin{split}
    \lim_{\nu\to\infty}\Req_+ 
    &= \frac{\gamma + \mu + \omega \kappa}{2 \beta \omega \kappa} \left[ 
    \left( 
        2 c_0 - 
        \frac{1}{\gamma + \mu}
    \right) c_1 -
    \frac{\abs{c_1}}{
         (\gamma + \mu)}
     \right].
\end{split}
\]

Here, we consider the behaviour of the equilibria as $\alpha\to 1^{+}$  and  $\alpha\to\infty$ 
    \[  \lim_{\alpha \to \infty} \Ieq_+= \frac{1}{\beta(\gamma+\mu)} \left[c_1+\frac{\gamma\kappa(\beta-(\gamma+\mu))}{\gamma+\mu+\kappa}\right],\]
  As a remark, the limit as $\alpha\to\infty$ and as $\alpha\to 1^+$ are the same.  
   \[ 
   \lim_{\alpha\to\infty}\Req_+
  =0,
\]
and lastly,

\[  \lim_{\alpha \to 1^+} \Req_+= 
    \frac{1}{\beta(\gamma+\mu)} \left[\frac{c_1\gamma}{\mu}+\frac{\gamma\kappa(\beta-(\gamma+\mu))}{\gamma+\mu+\kappa}\right].\]
    
\subsection{Numerical values of marked bifurcation points}
\label{app:numerical-values}

\begin{table}[H]
    \tbl{Critical points on the contour $y_\nu(\alpha) = 0$
    as marked in Figure~\ref{fig:RH-bubbles} and critical points on the limit cycles branch as marked in Figure~\ref{fig:greencurve}.}
    {\centering
    \begin{tabular}{@{}lrr@{}} \toprule
         ~ \qquad \qquad \qquad ~ 
         &\qquad ~ $\alpha$
         &\qquad ~ $\nu$\\
         \midrule
         $p_1 = (\alpha^*_1, \nu^*_1)$ 
         & $1.864273655292$ 
         & $2.063612920385$\\
         \midrule
         $p_2 = (\alpha^*_2, \nu^*_2)$
         & $2.0$ 
         & $2.063623848262$\\
         \midrule
         $p_3 = (\alpha^*_3, \nu^*_3)$ 
         & $2.157040937065$ 
         & $\nu^*_1$\\ 
         \midrule
         \midrule
         $p_4 = (\alpha^*_4, \nu^*_4)$ 
         & $1.366092512212$ 
         & $13.80272643151$\\
         \midrule
         $p_5 = (\alpha^*_5, \nu^*_5)$
         & $2.0$ 
         & $13.61692960743$\\
         \midrule
         $p_6 = (\alpha^*_6, \nu^*_6)$ 
         & $3.731549995264$ 
         & $\nu^*_4$\\ 
         \midrule
         \midrule
         $p_7 = (\alpha^*_7, \nu^*_7)$ 
         & $1.5987662507$ 
         & $14.9610290034$\\
         \midrule
         $p_8 = (\alpha^*_8, \nu^*_8)$ 
         & $2$ 
         & $14.936830813$\\
         \midrule
         $p_9 = (\alpha^*_9, \nu^*_9)$ 
         & $2.670631735$ 
         & $\nu^*_7$\\
         \bottomrule
    \end{tabular}}
    \label{app:tab-critical-points}
\end{table}

\begin{table}[H]
    \tbl{Critical GH points 
    as marked in Figure~\ref{fig:bif-diagram-large-alpha-nu}.}
    {\centering
    \begin{tabular}{@{}lrr@{}} \toprule
         ~ \qquad \qquad \qquad ~ 
         &\qquad ~ $\alpha$
         &\qquad ~ $\nu$\\
         \midrule
         $\gh_1 = (\alpha^*_{\gh_1}, \nu^*_{\gh_1})$ 
        & 1.1430260422
         &12.469198884\\
         \midrule
         $\gh_2 = (\alpha^*_{\gh_2}, \nu^*_{\gh_2})$
         & 7.9917337529 
         & 12.469198884\\
         \bottomrule
    \end{tabular}}
    \label{app:tab-critical-gh-points}
\end{table}

\end{document}